\documentclass{AIMS}
\usepackage{amsmath}
\usepackage{mathrsfs}
\usepackage{paralist}
\usepackage{epsfig}
\usepackage{float}  
  
\usepackage[colorlinks=true]{hyperref}
\usepackage{graphicx}
\usepackage{caption}
\usepackage{subcaption}
\usepackage{epstopdf}
\usepackage{newcmd}

\hypersetup{urlcolor=blue, citecolor=red}

\def\currentvolume{X}
 \def\currentissue{X}
  \def\currentyear{200X}
   \def\currentmonth{XX}
    \def\ppages{X--XX}

\newtheorem{theorem}{Theorem}[section]

\newtheorem{lemma}[theorem]{Lemma}
\newtheorem{proposition}{Proposition}

\theoremstyle{definition}

\newtheorem{remark}{Remark}

\title[multi-rate attitude estimation] 
      {Asymptotically Stable Optimal Multi-rate Rigid Body Attitude Estimation based on Lagrange-d'Alembert Principle}

\author[Maulik Bhatt, Amit K. Sanyal and Srikant Sukumar]{}


 \email{mcbhatt2@illinois.edu}
 \email{aksanyal@syr.edu}
 \email{srikant@sc.iitb.ac.in}

\thanks{$^*$ Corresponding author: Maulik Bhatt}

\begin{document}
\maketitle

\centerline{\scshape Maulik Bhatt $^*$}
\medskip
{\footnotesize
 \centerline{Systems and Control Engineering}
   \centerline{Indian Institute of Technology Bombay, 400076, India.}
} 

\medskip

\centerline{\scshape Amit K. Sanyal}
\medskip
{\footnotesize
 \centerline{Mechanical and Aerospace Engineering,}
   \centerline{Syracuse University, Syracuse, NY, USA.}
}

\medskip

\centerline{\scshape Srikant Sukumar}
\medskip
{\footnotesize
 \centerline{Systems and Control Engineering}
   \centerline{Indian Institute of Technology Bombay, 400076, India.}
}

\bigskip


\begin{abstract}
The rigid body attitude estimation problem is treated using the discrete-time Lagrange-d'Alembert principle. Three different possibilities are considered for the multi-rate relation between angular velocity measurements and direction vector measurements for attitude: 1) integer relation between sampling rates, 2) time-varying sampling rates, 3) non-integer relation between sampling rates. In all cases, it is assumed that angular velocity measurements are sampled at a higher rate compared to the inertial vectors. The attitude determination problem from two or more vector measurements in the body-fixed frame is formulated as Wahba's problem. At instants when direction vector measurements are absent, a discrete-time model for attitude kinematics is used to propagate past measurements. A discrete-time Lagrangian is constructed as the difference between a kinetic energy-like term that is quadratic in the angular velocity estimation error and an artificial potential energy-like term obtained from Wahba's cost function. An additional dissipation term is introduced and the discrete-time Lagrange-d'Alembert principle is applied to the Lagrangian with this dissipation to obtain an optimal filtering scheme. A discrete-time Lyapunov analysis is carried out to show that the optimal filtering scheme is asymptotically stable in the absence of measurement noise and the domain of convergence is almost global. For a realistic evaluation of the scheme, numerical experiments are conducted with inputs corrupted by bounded measurement noise. These numerical simulations exhibit convergence of the estimated states to a bounded neighborhood of the actual states.
\end{abstract}

\section{Introduction}\label{sec:1}
Spacecraft, underwater vehicles, aerial vehicles, and mobile robots require accurate knowledge of their orientation with respect to a known inertial frame. Typically, attitude estimators rely on the measurements of angular velocity and known inertial vectors in the body-fixed frame. Therefore, the rigid body attitude estimation problem using angular velocity and inertial vectors measurements in a body-fixed frame has been widely studied in past research. In practice, angular velocity is measured at a higher rate than inertial vectors. In this work\footnote{A preliminary arxiv version of this article is available at \cite{bhatt2020optimal}}, we address the attitude estimation problem given multi-rate measurements of inertially fixed vectors and angular velocity by minimizing the ``energy" stored in the state estimation errors.

One of the earliest solutions to attitude determination from vector measurements is the TRIAD algorithm, used to determine the rotation matrix from two linearly independent inertial vector measurements~\cite{black1964passive}. In \cite{wahba1965least}, Wahba presented the attitude determination problem as an optimization problem using three or more vector measurements where the cost function is the sum of the squared norms of vector errors. Various methods have been proposed in the literature to solve the Wahba's problem. Davenport was the first to reduce the Wahba's problem to finding the largest eigenvalue and the corresponding eigenvector of the so-called Davenport's K-matrix \cite{davenport1968vector}. In a similar approach, Mortari presented the EStimator of the Optimal Quaternion (ESOQ) algorithm in\cite{mortari1997esoq}, which provides the closed-form expressions of a $4\times4$ matrix's eigenvalues and then computes the eigenvector associated with the greatest of them, representing the optimal quaternion. The QUEST algorithm of \cite{shuster1981three} determines the attitude that achieves the best-weighted overlap of an arbitrary number of reference vectors. A Singular Value Decomposition (SVD) based method of solving the Wahba's problem was proposed by \cite{markley1988attitude}. Markley also devised a Fast Optimal Matrix Algorithm (FOMA) to solve Wahba's problem in \cite{markley1993attitude}. A coordinate-free framework of geometric mechanics was used in \cite{sany_acc06} to obtain a solution to Wahba's problem with robustness to measurement noise.  \cite{psiaki2012numerical} provides a numerical solution to Wahba's problem.

Attitude estimation methods based on minimizing ``energy" stored in state estimation errors can be found in \cite{zamani2010near,zamani2013minimum,izadi2014rigid, bhatt2020rigid}. Prior research that has designed attitude estimation schemes based on geometric mechanics includes \cite{mahony2008nonlinear,vasconcelos2007landmark,vasconcelos2008nonlinear,valpiani2008nonlinear}. Comprehensive surveys of various attitude estimation methods are available in \cite{crassidis2007survey,madinehi2013rigid}. However, most of the aforementioned schemes for attitude estimation work only in continuous time or measurement rich environments and neglect the sparsity in the inertial vector measurements. 
Inertial vector measurements are usually obtained with the help of Sun (and star) sensors or magnetometers which are accurate but suffer from lower sampling rates. \cite{sanyal2012attitude} provides one of the first solutions to rigid body attitude estimation with multi-rate measurements using uncertainty ellipsoids. A recursive method based on the cascade combination of an output predictor and an attitude observer can be found in \cite{khosravian2015recursive}. An attitude estimation scheme on the Special Orthogonal group using intermittent body-frame vector measurements was presented in \cite{BERKANE2019415}. In \cite{izadi2014rigid}, a filtering scheme in continuous-time is proposed by applying the Lagrange-d'Alembert principle on suitably formulated artificial kinetic and potential energy functions where the authors formulate filter equations assuming that inertial vector measurements and angular velocity measurements are available synchronously and continuously. A discrete-time estimation scheme in the presence of multi-rate measurements is proposed in \cite{bhatt2020rigid}. This work presents the Lyapunov analysis but does not contain a variational interpretation. 

The previous works in attitude estimations are presented in a continuous time domain and then discretized for numerical implementation. This voids the theoretical guarantee of asymptotic stability provided by the continuous-time estimator. Furthermore, most of the previous work in attitude estimation do not consider the realistic scenario when the inertial vectors and angular velocities are measured at a different rate. In the current work, we try to overcome these shortcoming by focusing on developing an asymptotically stable optimal geometric discrete-time attitude estimator based on the minimization of ``energy" stored in the errors of the state estimators in the presence of multi-rate measurements.
The measurements can be corrupted by noise and we do not assume any specific statistics (like normal distribution) on the measurement noise. However, the noise is assumed to be bounded. We represent the attitude as a rotation matrix which precludes potential singularity issues due to local coordinates (such as Euler angles). The multi-rate discrete-time filtering scheme presented here is obtained by applying the discrete-time Lagrange-D'Alembert principle \cite{marsden2001discrete} on a discrete-time lagrangian followed by a discrete-time Lyapunov analysis using a Lyapunov candidate that depends on the state estimation errors. The filtering scheme provided is asymptotically stable with an almost global region of convergence.

This paper is organized as follows. In the section \ref{sec:2}, the attitude estimation problem is formulated as Wahba's optimization problem and then some important properties of the Wahba's cost function are presented. In the section \ref{sec:3}, continuous-time rigid body attitude kinematics has been discretized and the propagation model for the measurements in the multi-rate measurement case is presented. Section \ref{sec:4} contains the application of variational mechanics to obtain a filter equation for attitude estimation. The filter equations obtained in the \ref{sec:4} are proven to be asymptotically stable with an almost global domain of convergence by deriving an appropriate dissipation torque using the discrete-time Lyapunov method in the \ref{sec:5}. Filter equations are numerically verified with realistic measurements (corrupted by bounded noise) in the \ref{sec:6}. Finally, \ref{sec:7} presents the concluding remarks with contributions and future work.

\section{Problem formulation and Notation}\label{sec:2}

\subsection{Notation and Preliminaries}

We define the trace inner product on $\bR^{m\times n}$ as
\begin{equation}
     \langle A_1,A_2\rangle := \text{trace}(A_1\T A_2).
\end{equation}
The group of orthogonal frame transformations on $\bR^3$ is defined by $\mathrm{O}(3) := \{Q \in \bR^{3\times 3} \; | \; \text{det}(Q)=\pm 1\}$. The Special orthogonal group on $\bR^3$ is denoted as $\SO$ defined as $\SO := \{R \in \bR^{3\times 3} \; | \; R\T R = RR\T = I_{3}\}$. The corresponding Lie algebra is denoted as $\mathfrak{so}(3) := \{ M \in \bR^{3\times 3} \; | \; M + M\T = 0\}$. Let $(\cdot)^\times:\bR^3 \rightarrow \mathfrak{so}(3) \subset \bR^{3\times3}$ be the skew-symmetric matrix cross-product operator and denotes the vector space isomorphism between $\bR^3$ and $\mathfrak{so}(3)$:
\begin{equation}
    v^\times = {\begin{bmatrix} v_1 \\ v_2 \\ v_3 \end{bmatrix}}^{\times} = \begin{bmatrix} 0 & -v_3 & v_2 \\ v_3 & 0 & -v_1\\ -v_2 & v_1 & 0 \end{bmatrix}.
\end{equation}
Further, let $\text{vex}(\cdot):\mathfrak{so}(3)\rightarrow\bR^3$ be the inverse of $(\cdot)^\times$. $\exp{(\cdot)}:\mathfrak{so}(3)\rightarrow \SO$ is the map defined as
\begin{equation}
        \exp{(M)} = \sum_{i=0}^{\infty}\frac{1}{k!}M^k.
\end{equation}
We define $Ad:\SO\times\mathfrak{so}(3)\rightarrow\mathfrak{so}(3)$ as
\begin{equation}
    Ad_R\Omega^\times = R\Omega^\times R\T = (R\Omega)^\times.
\end{equation}
In the rest of the article, the text ``Consider the time interval $[t_0,T] \subseteq \bR^+$'', indicates that the estimation process will be carried out for the time interval $[t_0,T]$ and the time interval is divided into $N$ equal sub-intervals $[t_i,t_{i+1}]$ for $i = 0,1,\dots,N$ with $t_N = T$ with time step size is denoted as, $h := t_{i+1} - t_i$; unless specified otherwise. {\color{black} Furthermore, throughout this article we use $(\cdot)^m$ to denote measured quantities. For example, if $\Omega$ is used to denote angular velocity, $\Omega^m$ denotes measured value of the angular velocity.}

\subsection{Attitude determination from vector measurements}

For the attitude estimation, we consider $k \in \mathbb{N}$ known and linearly independent inertial vectors in the body-fixed frame. Let's denote these vectors in the body-fixed frame by $u_j^m \in \bR^3$ for $j = 1,\ldots,k$, where $k \geq 2$. Note, that $k \geq 2$ is necessary for determining the attitude uniquely. When $k=2$, the cross product of the two measured vectors is used as the third independent measurement. Let $e_j\in \bR^3$ be the corresponding known inertial vectors. We denote the true vectors in the body-fixed frame by $u_j := R\T e_j$, where $R$ is the rotation matrix of the body-fixed frame with respect to the inertial frame. This rotation matrix provides a coordinate-free global and unique description of the attitude of the rigid body. Define the matrix populated by all $k$ measured vectors expressed in the body-fixed frame as column vectors as
\begin{align}\label{eq:1}
    U^m & = [\begin{matrix} u_1^{m} & u_2^{m} & u_1^{m}\times u_2^{m}\end{matrix}] \in \bR^{3\times3} \; \text{when} \; k=2 \; \text{and}, \nonumber\\
U^m & = [\begin{matrix} u_1^{m} & u_2^{m} & \ldots &u_k^{m}\end{matrix}] \in \bR^{3\times k} \; \text{when} \; k > 2,
\end{align}
and the corresponding inertial frame vectors as
\begin{align}\label{eq:2}
    E & = [\begin{matrix} e_1 & e_2 & e_1\times e_2\end{matrix}] \in \bR^{3\times3} \; \text{when} \; k=2 \;\text{and}, \nonumber\\
E & = [\begin{matrix} e_1 & e_2 & \ldots &e_k\end{matrix}] \in \bR^{3\times k} \; \text{when} \; k > 2.
\end{align}
The \emph{true} body vector matrix is as below:
\begin{align}\label{eq:3}
    U &= R^{T}E = [\begin{matrix} u_1 & u_2 & u_1\times u_2\end{matrix}] \in \bR^{3\times3} \; \text{when} \;  k=2 \; \text{and}, \nonumber\\
U & = R^{T}E = [\begin{matrix} u_1 & u_2 & \ldots &u_k\end{matrix}] \in \bR^{3\times k} \; \text{when} \; k > 2.
\end{align}

\subsubsection{Formulation of Wahba's cost function for instantaneous attitude determination from vector measurements}

The optimal attitude determination problem using a set of vector measurements is finding an estimated rotation matrix $\hat{R} \in \SO$,
such that the weighted sum of squared norms of the vector errors
\begin{equation}
    s_j = e_j - \hat{R}u_j^m
\end{equation}
is minimized. This attitude determination problem is known as Wahba's problem and consists of minimizing
\begin{equation}\label{eq:5}
     \mathcal{U}(\hat{R},U^{m}) = \frac{1}{2}\sum_{j=1}^{k}w_j (e_j - \hat{R}u_j^m)\T(e_j - \hat{R}u_j^m)
\end{equation}
with the respect to $\hat{R} \in \SO$, and the weights $w_j > 0$ for all $j \in \{1,2,\ldots,k\}$. we can express \eqref{eq:5} as
\begin{equation}\label{eq:6}
    \mathcal{U}(\hat{R},U^{m}) = \frac{1}{2} \langle\,E - \hat{R}U^m,(E-\hat{R}U^m)W\rangle,
\end{equation}
where $U^m$ is given by \eqref{eq:1}, $E$ is given by \eqref{eq:2}, and $W = \text{diag}(w_i)$ is the positive definite diagonal matrix of the weight factors for the measured directions. $W$ in \eqref{eq:6} can be generalized to be any positive definite matrix. It is to be noted that, having different weights for different vector measurements is helpful as in general we can have some sensors that provide more accurate measurements than others. Therefore, it makes sense to provide more weight to those accurate measurements during attitude estimation.

\subsubsection{Properties of Wahba's cost function in the absence of measurements errors}

{\color{black}This is an observer design that, like any other observer, can filter measurement noise from realistic measurements. As observer designs are based on perfect (deterministic) measurements for which their stability properties apply, it is useful to confirm these properties through numerical simulations.} We have $U^m = U = R\T E$  in the absence of measurement errors or noise. Let $Q = R\hat{R}\T \in \SO$ denote the attitude estimation error. The following lemmas from \cite{izadi2014rigid} stated here without proof give the structure and characterization of critical points of the Wahba's cost function.

\begin{lemma}\label{lemma:1}
Let rank($E$) = 3 and the singular value decomposition of $E$ be given by
\begin{align}
    & E := U_E\Sigma_EV_E\T \; \text{where} \; U_E \in O(3), V_E \in SO(m). \nonumber\\
 & \Sigma_E \in  Diag^+(3,m),
\end{align}
and Diag$^+(n_1,n_2)$ is the vector space of $n_1\times n_2$ matrices with positive entries along the main diagonal and all the other components zero. Let $\sigma_1,\sigma_2,\sigma_3$ denote the main diagonal entries of $\Sigma_E$. Further, Let $W$ from \eqref{eq:6} be given by
\begin{equation}
    W = V_EW_0V_E\T \; \text{where} \; W_0 \in \text{Diag}^+(m,m)
\end{equation}
and the first three diagonal entries of $W_0$ are given by
\begin{equation}
    w_1 = \frac{d_1}{\sigma_1^2}, \;  w_2 = \frac{d_2}{\sigma_2^2}, \;  w_3 = \frac{d_3}{\sigma_3^2} \;\; \text{where} \; d_1,d_2,d_3 > 0.
\end{equation}
Then, $K = EWE\T$ is positive definite and
\begin{equation}\label{eq:10}
    K = U_E\Delta U_E\T \; \text{where} \; \Delta = \text{diag}(d_1,d_2,d_3)
\end{equation}
is its eigen decomposition. Moreover, if $d_i \neq d_j$ for $i \neq j$ and $i,j \in \{1,2,3\}$ then $\langle\,I-Q,K\rangle$ is a Morse function whose set of critical points given as the solution of $S_K(Q) := \text{vex}\left(KQ - Q\T K\right) = 0$:
\begin{equation}\label{eq:11}
    C_Q := \{I,Q^1,Q^2,Q^3\} \; \text{where} \; Q^i = 2U_Ea_ia_i\T U_E\T - I
\end{equation}
and $a_i$ is the $i^{th}$ column vector of the identity matrix $I \in \SO$.
\end{lemma}

\begin{lemma}\label{lemma:2}
Let $K =EWE\T$ have the properties given by lemma \ref{lemma:1}. Then the map $\SO \ni Q  \mapsto \langle\,I-Q,K\rangle \in \bR$  with critical points given by \eqref{eq:11} has a global minimum at the identity $I\in \SO$, a global maximum and two hyperbolic saddle points whose indices depend on the distinct eigenvalues $d_1,d_2,$ and $d_3$ of $K$. 
\end{lemma}

\section{Discretization of Attitude Kinematics}\label{sec:3}

\subsection{\bf Case-1: The sampling rates are integer related}

Consider the time interval $[t_0,T] \subseteq \bR^+$. Let the true angular velocity in the body-fixed frame be denoted by $\Omega \in \bR^3$. The true and measured angular velocities at the time instant $t_i$ will be denoted by $\Omega_i$ and $\Omega_i^m$ respectively. Further, let $U_i$ and $U_{i}^m$ denote the matrix formed by true and measured inertial vectors in the body-fixed frame at the time instant $t_i$ respectively. The assumption is that angular velocity measurements and inertial vectors measurements in the body-fixed frame are coming at a different but constant rate. In general coarse rate gyros have much higher sampling rate than that of a coarse attitude sensor. Therefore, in a realistic scenario, angular velocities are measured at a higher rate than the inertial vector measurements in the body-fixed frame. Therefore, we assume that the measurements of angular velocity ($\Omega^m$) are available after each time interval $h$ say, $\Omega^m_0,\Omega^m_1,\dots,\Omega^m_N$ while, inertial vector measurements in the body-fixed frame are available after time interval $nh, n \in N$ say, $U^m_0,U^m_n,U^m_{2n},\dots$. \\

We have, $U = R\T E$. Therefore, at time instants $t_i$ and $t_{i+1}$, the following relations will hold true respectively; $U_i = R_i\T E_i, \; U_{i+1} = R_{i+1}\T E_{i+1}$. Here, $R_i$ and $R_{i+1}$ are the rotation matrices from body-fixed frame to inertial frame at time instants $t_i$ and $t_{i+1}$ respectively. $E_i = E_{i+1} = E$ are the corresponding known vectors expressed in the inertial frame. Note that the vectors are fixed in the inertial frame and do not change with the time. The continuous time attitude kinematics are,
\begin{equation}\label{eq:12}
    \Dot{R} = R\Omega^\times.
\end{equation}
We discretize the kinematics in \eqref{eq:12} as follows,
\begin{equation}\label{eq:13}
    R_{i+1} = R_i\exp{\left(\frac{h}{2}(\Omega_{i+1}+\Omega_i)^\times\right)},
\end{equation}
where, $\exp{(\cdot)}:\mathfrak{so}(3)\mapsto \SO$ is the map defined as,
\begin{equation}
    \exp{(M)} = \sum_{k=0}^{\infty}\frac{1}{k!}M^k.
\end{equation}

Using \eqref{eq:3} and the discretization from \eqref{eq:13} we get,
\begin{align}\label{eq:15}
    U_{i+1} & = \exp{\left(-\frac{h}{2}(\Omega_{i+1}+\Omega_i)^\times\right)}R_i\T E_i \nonumber\\ & = \exp{\left(-\frac{h}{2}(\Omega_{i+1}+\Omega_i)^\times\right)}U_i.
\end{align}

For the instants of time when inertial vector measurements in the body-fixed frame are not available we will use \eqref{eq:15} to obtain the missing values of $U_i^m$. This implies that for the time instants $ (n-1)h < t_i < nh, n \in \mathbb{N}$, by employing the propagation scheme in \eqref{eq:15}, we propagate direction vector measurements between the instants at which they are measured, using the angular velocity measurements that are obtained at a faster rate. We now formalise the aforementioned inertial vector measurement model as below,
\begin{equation}\label{eq:16}
    \Tilde{U}_i^m := \begin{cases}
        U_i^m, & \text{if} \; i\mod n = 0
        \\
        \exp{\left(-\frac{h}{2}(\Omega_{i-1}^m+\Omega_i^m)^\times\right)}\Tilde{U}^m_{i-1}, & \text{otherwise}.
    \end{cases}
\end{equation}

Note that in the absence of measurements errors, we have $\Omega_i^m = \Omega_i,\, \forall i \in \{0,1,\ldots,N\}$.  Also, $U_i^m = U_i$ for the time instants when inertial vector measurements are available. Now, at time instant $t_0$, we have $\Tilde{U}^m_{0}=U^m_0 = U_0$ and $\Omega^m_0 = \Omega_0$. Using \eqref{eq:16} at time  instant $t_1$, noting that $\Omega^m_1 = \Omega_1$, we get $\Tilde{U}^m_{1} = \exp{\left(-\frac{h}{2}(\Omega_{0}+\Omega_1)^\times\right)}{U}_{0}$. Comparing it with \eqref{eq:15}, we have $\Tilde{U}^m_{1} = U_1$. Using the relation from \eqref{eq:3} we have $\Tilde{U}^m_{1} = R_1\T E_1$. Similarly, combining \eqref{eq:15}, and \eqref{eq:16}, and using the relation in \eqref{eq:3} we get the following relation for all $i \in \{0,1,\ldots,N\}$ in the absence of measurement errors:
\begin{equation}
    \Tilde{U}_i^m = R_i\T E_i.
\end{equation}

\subsection{\bf Case-2: The sampling rates are time-varying}

Consider the time interval $[t_0,T] \subseteq \bR^+$. Let the true angular velocity in the body-fixed frame be denoted by $\Omega \in \bR^3$. The true and measured angular velocities at the time instant $t_i$ will be denoted by $\Omega_i$ and $\Omega_i^m$ respectively. It is assumed that a measurement $U^m_0$ is available at time instant $t_0$. Now the next measurement will be available after a time interval $nh$ where $n$ is an integer variable taking values between $n_1$ and $n_2$ which are user provided. Note that since $n$ is varying, that sampling rate of inertial vector sensor is not fixed but varying over time. Considering the discrete kinematics of $R$ and $U$ from \eqref{eq:13} and \eqref{eq:15} we propose the following discrete-time vector measurement model,
\begin{equation}
    \Tilde{U}_i^m := \begin{cases}
        U_i^m, & \text{if} \; \text{measurement available}
        \\
        \exp{\left(-\frac{h}{2}(\Omega_{i-1}^m+\Omega_i^m)^\times\right)}\Tilde{U}^m_{i-1}, & \text{otherwise}.
    \end{cases}
\end{equation}
Note that now we have obtained values of $\Omega^m_i$ and $\Tilde{U}^m_i$ for each time instant $t_i$ and similar to the previous arguments, it can be proven that in the absence of measurement errors, we have $\Omega^m_i = \Omega_i$ and $\Tilde{U}_i^m = R_i\T E_i$.

\subsection{\bf Case-3: The sampling rates are non-integer related}

Again consider the time interval $[t_0,T] \subseteq \bR^+$. Let the angular velocity measurements $\Omega^m \in \bR^3$ be available after each time interval of $h$. Now suppose the inertial vector measurements are available after a time interval $rh$ where $r = n_1/n_2$ is a rational number with $n_1,n_2 \in \mathbb{N}$. Let $n$ be the greatest common factor of $n_1$ and $n_2$. Let $h^\prime = nh/n_2$. Now, consider the time interval $[t_0,T] \subseteq \bR^+$ divided into $N^\prime$ equal sub-intervals $[t^\prime_i,t^\prime_{i+1}]$ for $i = 0,1,\dots,N^\prime$ with $t^\prime_N = T $ and let $t^\prime_{i+1} - t^\prime_i = h^\prime$. Note that a measurement $\Omega^m$ will be available after a time interval of $n_2h^\prime/n$ and a measurement $U^m$ will be available after time interval $rh = \frac{n_1}{n_2}\times\frac{n_2}{n}h^\prime = n_1h^\prime/n$. \\

We define the measurement model for $\Omega^m$ as
\begin{equation}
    \Tilde{\Omega}_i^m := \begin{cases}
        \Omega_i^m, & \text{if} \; i \mod n_2/n = 0
        \\
        \Omega_{i-1}^m, & \text{otherwise}.
    \end{cases}
\end{equation}
for all $i = 0,1,\dots,N^\prime$. Now we can define the vector measurement model to be,
\begin{equation}
    \Tilde{U}_i^m := \begin{cases}
        U_i^m, & \text{if} \; i \mod n_1/n = 0
        \\
        \exp{\left(-\frac{h}{2}(\Omega_{i-1}^m+\Omega_i^m)^\times\right)}\Tilde{U}^m_{i-1}, & \text{otherwise}.
    \end{cases}
\end{equation}

Note that the division of the time interval $[t_0, T]$ with the time step size of $h^\prime$ is finer than that of with the time step size of $h$ with the relation as $h = \frac{n_2}{n}h^\prime$. Therefore for each $i \in \{0,1,\ldots,N\}$ we have $j = \frac{n_2}{n}i \in \{0,1,\ldots,n^\prime\}$ such that $t_i = t_j^\prime$. Redefining $\Omega^m_i :=  \Tilde{\Omega}^m_{\frac{n_2}{n}i}$, and $\Tilde{U}^m_i = \Tilde{U}^m_{\frac{n_2}{n}i}$ for all $i \in \{0,1,\ldots,N\}$. Note that now we have obtained values of $\Omega^m_i$ and $\Tilde{U}^m_i$ for each time instant $t_i$ and similar to previous arguments, it can be proven that in the absence of measurement errors, we have $\Omega^m_i = \Omega_i$ and $\Tilde{U}_i^m = R_i\T E_i$. \\

Observe that in any of the aforementioned three cases, the measurement models provide values $\Omega^m_i$ and $\Tilde{U}^m_i$ for all the discrete-time instants $t_i, \; i \in \{0,1,\ldots,N\}$ in the time interval $[t_0, T]$.

\section{Discrete-time optimal attitude estimator based on Lagrange-d'Alembert's principle}\label{sec:4}

The value of the Wahba's cost function at each instant encapsulates the error in the attitude estimation. We can consider the Wahba's cost function as an artificial potential energy-like term. Therefore using \eqref{eq:6} we have
\begin{equation}\label{eq:17}
    \mathcal{U}_i = \mathcal{U}(\hat{R}_i,\Tilde{U}^{m}_i) = \frac{1}{2} \langle\,E_i - \hat{R}_i\tilde{U}^m_i,(E_i-\hat{R}_i\Tilde{U}_i^m)W_i\rangle,
\end{equation}
where $\Tilde{U}_i^m$ is according to the inertial vector propagation model presented in section \ref{sec:3}. The term encapsulating the ``energy" in the angular velocity estimation error is denoted by the map $\mathcal{T}^v$ defined as
\begin{align} \label{eq:18}
    & \mathcal{T}^v_i := \mathcal{T}^v (\hat{\Omega}_i,\Omega^m_i, \hat{\Omega}_{i+1},\Omega^m_{i+1}) := \nonumber\\ &\frac{m}{2}(\Omega^m_i + \Omega^m_{i+1} - \hat{\Omega}_i - \hat{\Omega}_{i+1})\T(\Omega^m_i + \Omega^m_{i+1} - \hat{\Omega}_i - \hat{\Omega}_{i+1}),
\end{align}
where $m>0$ is a scalar and $\hat{\Omega}_i$ stands for the value of estimated angular velocity at the discrete time instant $t_i$. We can write \eqref{eq:18} in terms of the angular velocity estimation error $\omega_i := \Omega^m_i - \hat{\Omega}_i$ as
\begin{equation}\label{eq:19}
    \mathcal{T}^v (\omega_i,\omega_{i+1}) = \frac{m}{2}(\omega_i + \omega_{i+1})\T(\omega_i + \omega_{i+1}).
\end{equation}
The discrete time Lagrangian can be written as:
\begin{align}\label{eq:20}
& \mathscr{L}(\hat{R}_i,\Tilde{U}_i^m,\omega_i,\omega_{i+1}) = \mathcal{T}^v(\omega_i,\omega_{i+1}) - \mathcal{U}(\hat{R}_i,\Tilde{U}^{m}_i) \nonumber \\ & \qquad =   \frac{m}{2}(\omega_i + \omega_{i+1})\T(\omega_i + \omega_{i+1}) - \frac{1}{2} \langle\,E_i - \hat{R}_i\tilde{U}^m_i,(E_i-\hat{R}_i\Tilde{U}_i^m)W_i\rangle. 
\end{align}
We use variational mechanics \cite{goldstein1980classical,greenwood1997classical} approach to obtain an optimal estimation scheme for the constructed Lagrangian. In the absence of a dissipative term, the variational approach would result in (a generalization to the Lie group of) the Euler-Lagrange equations that we obtain in the context of optimal control. The generalization would be an Euler-Poincare equation on \SO~\cite{blbook}. Therefore, if the estimation process is started at time $t_0$, then the discrete-time action functional corresponding to the discrete-time Lagrangian \eqref{eq:20} over the time interval $[t_0, T]$ can be expressed as
\begin{multline}\label{eq:21}
    \mathfrak{s}_d(\mathscr{L}(\hat{R}_i,\Tilde{U}_i^m,\omega_i,\omega_{i+1})) = h\sum_{i=0}^N \left(\mathscr{L}(\hat{R}_i,\Tilde{U}_i^m,\omega_i,\omega_{i+1})\right) \\
    = h\sum_{i=0}^N \bigg\{
    \frac{m}{2}(\omega_i + \omega_{i+1})\T(\omega_i + \omega_{i+1}) \bigg. \left. - \frac{1}{2} \langle\,E_i - \hat{R}_i\Tilde{U}^m_i,(E_i-\hat{R}_i\Tilde{U}^m_i)W_i\rangle \right\}.
\end{multline}

\subsection{Discrete-time attitude state estimation based on the discrete-time Lagrange-d'Alembert principle}

Consider attitude state estimation in discrete-time in the presence of multirate measurements with noise and initial state estimate errors. Applying the discrete-time Lagrange-d’Alembert principle \cite{marsden2001discrete} from variational mechanics to the action functional $\mathfrak{s}_d(\mathscr{L}(\hat{R}_i,\Tilde{U}_i^m,\omega_i,\omega_{i+1}))$ given by \eqref{eq:21}, in the presence of a dissipation term on $\omega_i = \Omega_i^m - \hat{\Omega}_i $, leads to the following attitude and angular velocity filtering scheme.

\begin{proposition}\label{prop}
Consider the time-interval $[t_0,T]$. {\color{black}Consider the multi-rate measurement models from Section \ref{sec:3} such that we have values $\Omega^m_i$ and $\Tilde{U}^m_i$ for all the discrete-time instants $t_i, \; i \in \{0,1,\ldots,N\}$ in the time interval $[t_0, T]$.} Let the $W_i$ be chosen such that $K_i = E_iW_iE_i\T$ satisfies eigen decomposition condition \eqref{eq:10} of lemma \ref{lemma:1}. Also, let $\tau_{D_i} \in \bR^3$ denote the value of the dissipation torque at the time instant $t_i$. A discrete-time optimal filter obtained by applying the discrete-time Lagrange-d’Alembert principle would be as follows:
\begin{equation}\label{eq:22}
    \begin{cases}
    \hat{R}_{i+1} = \hat{R}_i\exp{\left(\frac{h}{2}(\hat{\Omega}_{i+1}+\hat{\Omega}_i)^\times\right)} \\
    
    m(\omega_{i+2} + \omega_{i+1}) = \exp{\left(-\frac{h}{2}(\hat{\Omega}_{i+2}+\hat{\Omega}_{i+1})^\times\right)} \Big \{ m(\omega_{i+1}+\omega_i) \Big.
    \\ \left. \qquad \qquad \qquad \qquad +   \frac{h}{2}S_{L_{i+1}}(\hat{R}_{i+1}) - \frac{h}{2}\tau_{D_{i+1}} \right\}\\
    
    \hat{\Omega}_i = \Omega^m_i - \omega_i,
    \end{cases}
\end{equation}
where $S_{L_{i}}(\hat{R}_i) = \text{vex}(L_i\T\hat{R}_i - \hat{R}_i\T L_i) \in \bR^3$, $L_i = E_iW_i(\Tilde{U}^m_i)\T$, and $(\hat{R}_0,\hat{\Omega}_0) \in \SO\times\bR^{3\times3}$ are intial estimated states.
\end{proposition}
\begin{proof}
Consider a first variation in the discrete attitude estimate as
\begin{equation}\label{eq:23}
    \delta \hat{R}_i = \hat{R}_i\Sigma_i^{\times},
\end{equation}
where $\Sigma_i \in \bR^3$ represents a variation   for the discrete attitude estimate. For fixed end-point variations, we have $\Sigma_0 = \Sigma_N = 0$. A first order approximation is to assume that $\hat{\Omega}^\times$ and $\delta\hat{\Omega}^\times$ commute. Taking the first variation of the discrete-time attitude kinematics according to the first equation of \eqref{eq:22} and comparing with \eqref{eq:23} we get
\begin{align}\label{eq:24}
    \delta\hat{R}_{i+1} & = \delta\hat{R}_i\exp{\left(\frac{h}{2}(\hat{\Omega}_{i+1} + \hat{\Omega}_i)^\times\right)} + \frac{h}{2}\hat{R}_i\exp{\left( \frac{h}{2}(\hat{\Omega}_{i+1}+\hat{\Omega}_i)^\times \right)}\delta(\hat{\Omega}_{i+1}+\hat{\Omega}_i)^\times \nonumber \\
    & = \hat{R}_{i+1}\Sigma_{i+1}^\times.
\end{align}
\eqref{eq:24} can be rearranged to
\begin{align}\label{eq:25}
    \hat{R}_{i+1}\frac{h}{2}\delta(\hat{\Omega}_{i+1}+\hat{\Omega}_i)^\times & = \hat{R}_{i+1}\Sigma_{i+1}^\times - \hat{R}_{i+1}\text{Ad}_{\exp{\left(-\frac{h}{2}(\hat{\Omega}_{i+1}+\hat{\Omega}_i)^\times\right)}}\Sigma_i^\times \nonumber \\
    \Rightarrow \frac{h}{2}\delta(\hat{\Omega}_{i+1}+\hat{\Omega}_i)^\times & = \Sigma_{i+1}^\times - \text{Ad}_{\exp{\left(-\frac{h}{2}(\hat{\Omega}_{i+1}+\hat{\Omega}_i)^\times\right)}}\Sigma_i^\times.
\end{align}
\eqref{eq:25} can be equivalently written as an equation in $\bR^3$ as follows:
\begin{equation}\label{eq:26}
    \frac{h}{2}\delta(\hat{\Omega}_{i+1}+\hat{\Omega}_i) = \Sigma_{i+1} - \exp{\left(-\frac{h}{2}(\hat{\Omega}_{i+1}+\hat{\Omega}_i)^\times\right)}\Sigma_i.
\end{equation}
Note that, $\omega_i = \Omega_i^m - \hat{\Omega}_i$ gives us
\begin{equation}\label{eq:27}
    \delta(\omega_{i+1}+\omega_i) = - \delta(\hat{\Omega}_{i+1}+\hat{\Omega}_i).
\end{equation}
Consider artificial potential energy term as expressed in \eqref{eq:17}. Taking its first variation with the respect to estimated attitude $\hat{R}$, we get
\begin{align}\label{eq:28}
    \delta\mathcal{U}_i & = \frac{1}{2}\left\{ \langle -\delta\hat{R}_i\Tilde{U}_i^m , (E_i-\hat{R}_i\Tilde{U}_i^m)W_i \rangle \right. \left. + \langle E_i-\hat{R}_i\Tilde{U}_i^m , (-\delta\hat{R}_i\Tilde{U}_i^m)W_i \rangle \right\} \nonumber \\
    & = \langle -\delta\hat{R}_i\Tilde{U}_i^m , (E_i-\hat{R}_i\Tilde{U}_i^m)W_i \rangle = \langle -\hat{R}_i\Sigma_i^\times, (E_i-\hat{R}_i\Tilde{U}_i^m)W_i \rangle \nonumber \\
    & = \text{trace}\left( (\Tilde{U}_i^m)\T\Sigma_i^\times\hat{R}_i\T(E_i-\hat{R}_i\Tilde{U}_i^m)W_i \right) = \text{trace}\left((\Sigma_i^\times)\T\Tilde{U}_i^mW_iE_i\T\hat{R}_i\right) \nonumber \\
    &= \langle \Sigma_i^\times , \Tilde{U}_i^mW_iE_i\T\hat{R}_i\rangle = \frac{1}{2} \langle\, \Sigma^\times , \Tilde{U}_i^mW_iE_i\T\hat{R}_i - \hat{R}_i\T E_iW_i(\Tilde{U}_i^m)\T\rangle \nonumber \\
    &= \frac{1}{2}\langle\, \Sigma_i^\times , L_i\T\hat{R}_i - \hat{R}_i\T L_i \rangle = S_{L_i}\T(\hat{R}_i)\Sigma_i.
\end{align}
Consider the first variation in the artificial kinetic energy $\mathcal{T}^v(\omega_i,\omega_{i+1})$ as in \eqref{eq:19} with the respect to the angular velocity estimation error,
\begin{align}
    \delta\mathcal{T}^v_i = m(\omega_i + \omega_{i+1})\T\delta(\omega_i + \omega_{i+1}).
\end{align}
Using the results in \eqref{eq:26} and \eqref{eq:27} we get
\begin{align}\label{eq:30}
    \delta\mathcal{T}^v_i & = -m(\omega_i + \omega_{i+1})\T\delta(\Omega_i + \Omega_{i+1}) \nonumber \\
    & = \frac{2}{h}m(\omega_i + \omega_{i+1})\T\left(\exp{\left(-\frac{h}{2}(\hat{\Omega}_{i+1}+\hat{\Omega}_i)^\times\right)}\Sigma_i - \Sigma_{i+1}\right).
\end{align}
The first variation of the discrete-time action sum in \eqref{eq:21} using \eqref{eq:28} and \eqref{eq:30} can be written as
\begin{align}
    \delta\mathfrak{s}_d & = h\sum_{i=0}^N\left\{ \delta\mathcal{T}_i^v - \delta\mathcal{U}_i \right\} \nonumber \\
    & = \sum_{i=0}^N\left\{ 2m(\omega_i + \omega_{i+1})\T\exp{\left(-\frac{h}{2}(\hat{\Omega}_{i+1}+\hat{\Omega}_i)^\times\right)}\Sigma_i \right. \nonumber \\ & \qquad \bigg. - 2m(\omega_i + \omega_{i+1})\T\Sigma_{i+1} - hS_{L_i}\T(\hat{R}_i)\Sigma_i \bigg\}.
\end{align}
Applying the discrete-time  Lagrange-d'Alembert principle to the attitude motion, we obtain
\begin{align}\label{eq:32}
    \delta\mathfrak{s}_d + h\sum_{i=0}^{N-1} \tau_{D_i}\T\Sigma_i = 0 & \Rightarrow \sum_{i=0}^{N-1}\left\{ 2m(\omega_i + \omega_{i+1})\T\left[\exp{\left(-\frac{h}{2}(\hat{\Omega}_{i+1}+\hat{\Omega}_i)^\times\right)}\Sigma_i \right. \right. \nonumber \\ & \bigg. \bigg. \qquad - \Sigma_{i+1} \bigg]  - hS_{L_i}\T(\hat{R}_i)\Sigma_i + h\tau_{D_{i}}\T\Sigma_i \bigg\} = 0.
\end{align}
For $0 \leq i < N$, \eqref{eq:32} leads to
\begin{align}
     & 2m(\omega_{i+2} + \omega_{i+1})\T\exp{\left(-\frac{h}{2}(\hat{\Omega}_{i+2}+\hat{\Omega}_{i+1})^\times\right)} - \nonumber \\
     & 2m(\omega_{i+1} + \omega_i)\T - hS_{L_{i+1}}\T(\hat{R}_{i+1}) + h\tau_{D_{i+1}}\T = 0 \nonumber \\
     & \Rightarrow 2m\exp{\left(-\frac{h}{2}(\hat{\Omega}_{i+2}+\hat{\Omega}_{i+1})^\times\right)}(\omega_{i+2} + \omega_{i+1}) \nonumber \\ & =
      2m(\omega_{i+1} + \omega_i) + hS_{L_{i+1}}(\hat{R}_{i+1}) - h\tau_{D_{i+1}},
\end{align}
which in turn leads to the second filter equation.
\end{proof}

{\color{black}
\begin{remark}
Please note that, we introduced a virtual dissipation torque in the Proposition \ref{prop} and the performance of the estimator will depend on the value of dissipation torque. In the next section, we determine the value of dissipation torque via discrete-Lyapunov analysis so that the resulting estimator is almost globally asymptotically stable.
\end{remark}}

\section{Stability of the filter using the discrete Lyapunov Approach}\label{sec:5}

For the Lyapunov stability of the filter equations, we need to construct a suitable Lyapunov candidate function. We use the Wahba's cost function expressed in \eqref{eq:17} as the artificial potential energy which encapsulates the error in the estimation of attitude. A new term encapsulating the ``energy" in the angular velocity estimation error can be constructed as the map $\mathcal{T}^l:\bR^3\times\bR^3\rightarrow \bR$ defined as
\begin{equation}\label{eq:34}
    \mathcal{T}^l_i := \mathcal{T}^l (\hat{\Omega}_i,\Omega^m_i)  := \frac{m}{2}(\Omega^m_i - \hat{\Omega}_i)\T(\Omega^m_i - \hat{\Omega}_i),
\end{equation}
where $m>0$ is a scalar same as before. Further, \eqref{eq:34} can be written in terms of angular velocity estimation error, $\omega_i := \Omega^m_i - \hat{\Omega}_i$ as follows:
\begin{equation}
    \mathcal{T}^l (\omega_i) = \frac{m}{2}(\omega_i)\T(\omega_i).
\end{equation}
In the absence of measurement errors, we have $\Tilde{U}_i^m = R_i\T E_i$. Therefore we can we can write \eqref{eq:17} in terms of state estimation error $Q_i = R_i\hat{R}\T_i$ as
\begin{align}
     \mathcal{U}(\hat{R}_i,\Tilde{U}^{m}_i) & = \frac{1}{2} \langle\,E_i - \hat{R}_iR_i\T E_i,(E_i-\hat{R}_iR_i\T E_i)W_i\rangle \nonumber \\
     & = \langle\,I - R_i\hat{R}_i\T,E_iW_iE_i\T\rangle \nonumber \\
     \Rightarrow \mathcal{U}_i = \mathcal{U}(Q_i) & = \langle\,I-Q_i,K_i\rangle \; \text{where} \; K_i = E_iW_iE_i\T.
\end{align}
The weights $W_i$'s are chosen such that $K_i$ is always positive definite with distinct eigenvalues according to lemma \ref{lemma:1}.

\begin{theorem}
Consider the time-interval $[t_0,T]$. {\color{black} Consider the multi-rate measurement models from Section \ref{sec:3} such that we have values $\Omega^m_i$ and $\Tilde{U}^m_i$ for all the discrete-time instants $t_i, \; i \in \{0,1,\ldots,N\}$ in the time interval $[t_0, T]$.} Then the estimation scheme in Proposition \ref{prop} with the following value of the dissipation torque:
\begin{align}\label{eq:37}
    &\tau_{D_{i+1}} = \frac{1}{h}\bigg\{ 2m(\omega_{i+1} + \omega_i) + hS_{L_{i+1}}(\hat{R}_{i+1}) - \frac{2m}{m+l}\bigg. \nonumber \\ & \left. \exp{\left(\frac{h}{2}(\hat{\Omega}_{i+2}+\hat{\Omega}_{i+1})^\times\right)}\left[ 2m\omega_{i+1} + k_phS_{L_{i+1}}(\hat{R}_{i+1}) \right] \right\},
\end{align}
leads to the estimation scheme
\begin{equation}\label{eq:38}
    \begin{cases}
    \omega_{i+1} = \frac{1}{m+l}\left[ (m-l)\omega_i + k_phS_{L_i}(\hat{R}_i) \right]\\ 
    \hat{\Omega}_i = \Omega^m_i - \omega_i \\
    \hat{R}_{i+1} = \hat{R}_i\exp{\left(\frac{h}{2}(\hat{\Omega}_{i+1}+\hat{\Omega}_i)^\times\right)},
    \end{cases}
\end{equation}
where $S_{L_{i}}(\hat{R}_i) = \text{vex}(L_i\T\hat{R}_i - \hat{R}_i\T L_i) \in \bR^3$, $L_i = E_iW_i(\Tilde{U}^m_i)\T$, $l>0$, $l \neq m$ and $k_p > 0$, which is asymptotically stable at the estimation error state $(Q,\omega) := (I,0)$ ($Q_i = R_i\hat{R}\T_i$) in the absence of measurement errors. Further, the domain of attraction of $(I,0)$ is a dense open subset of $\SO\times\bR^3$.
\end{theorem}
\begin{proof}
Using the third equation from \eqref{eq:38} we have
\begin{align} \label{eq:39}
     Q_{i+1} & = R_{i+1}\hat{R}\T_{i+1} \nonumber \\ & = Q_i\hat{R}_i\exp{\left(\frac{h}{2}(\omega_{i+1} + \omega_i)^\times\right)}\hat{R}_i\T.
\end{align}
We choose the following discrete-time Lyapunov candidate:
\begin{equation}\label{eq:40}
    V_i := V(Q_i,\omega_i) := k_p\mathcal{U}_i + \mathcal{T}^l_i,
\end{equation}
where $k_p>0$ is a constant.

The stability of the attitude and angular velocity estimation error can be shown by analyzing $\Delta V_i = k_p\Delta\mathcal{U}_i + \Delta\mathcal{T}^l_i$.

Assuming $K_i$ to be constant and letting $K = K_i = K_{i+1}$ we obtain
\begin{align}\label{eq:41}
     & \Delta\mathcal{U}_i = \mathcal{U}_{i+1} - \mathcal{U}_i = \langle\,I-Q_{i+1},K\rangle - \langle\,I-Q_i,K\rangle \nonumber \\
     & \Delta\mathcal{U}_i = \langle\,Q_i - Q_{i+1},K\rangle = -\langle\,\Delta Q_i,K\rangle,
\end{align}
where $\Delta Q_i = Q_{i+1} - Q_i$. Therefore,
\begin{align}
    \Delta Q_i & = Q_{i+1} - Q_i \nonumber \\
    & = Q_i\left[\hat{R}_i\exp{\left(\frac{h}{2}(\omega_{i+1} + \omega_i)^\times\right)}\hat{R}_i\T - I\right].
\end{align}
Considering the first order expansion of  $\exp{\left(\frac{h}{2}(\hat{\omega}_{i+1} + \hat{\omega}_i)^\times\right)}$ as
\begin{equation}\label{eq:47}
    \exp{\left(\frac{h}{2}(\omega_{i+1} + \omega_i)^\times\right)} \approx I + \frac{h}{2}(\omega_{i+1} + \omega_i)^\times,
\end{equation}
we have
\begin{align}
    \Delta Q_i & = Q_i\left[\hat{R}_i\left(I + \frac{h}{2}(\omega_{i+1} + \omega_i)^\times\right)\hat{R}_i\T - I\right] \nonumber \\
    & = \frac{h}{2}Q_i\left(\hat{R}_i(\omega_{i+1} + \omega_i)^\times\hat{R}_i\T\right) \nonumber \\
    & = \frac{h}{2}Q_i\left(\hat{R}_i(\omega_{i+1} + \omega_i)\right)^\times .
\end{align}
It has to be noted that approximation in \eqref{eq:47} is accurate for small values of $h$ and may affect the stability results for very high values of $h$. In the absence of measurement errors, we have $\Tilde{U}_i^m = R_i\T E_i$. Therefore, it follows that
\begin{align}
    \Delta\mathcal{U}_i & = -\frac{h}{2}\left\langle\,Q_i\left( \hat{R}_i \left(\omega_{i+1} + \omega_i \right) \right)^\times,K\right\rangle \nonumber \\
    & = -\frac{h}{2}\left \langle\,R_i(\omega_{i+1} + \omega_i)^\times\hat{R}_i\T,E_iW_iE_i\T\right\rangle \nonumber\\
    & = -\frac{h}{2}\left \langle\,(\omega_{i+1} + \omega_i)^\times\hat{R}_i\T,R_i\T E_iW_iE_i\T\right\rangle \nonumber\\
    & = -\frac{h}{2}\left \langle\,(\omega_{i+1} + \omega_i)^\times\hat{R}_i\T,\Tilde{U}_i^mW_iE_i\T\right\rangle,
\end{align}
and noting that $L_i = E_iW_i(\Tilde{U}_i^m)\T$, we get
\begin{align}\label{eq:46}
    \Delta\mathcal{U}_i & = -\frac{h}{2}\left \langle\,(\omega_{i+1} + \omega_i)^\times,L_i\T\hat{R}_i\right\rangle \nonumber\\
    & = -\frac{h}{4}\left \langle\,(\omega_{i+1} + \omega_i)^\times,L_i\T\hat{R}_i - \hat{R}_i\T L_i\right\rangle \nonumber\\
    & = -\frac{h}{2}(\omega_{i+1}+\omega_i)\T S_{L_i}(\hat{R}_i).
\end{align} 
Similarly we can compute the change in the kinetic energy as follows:
\begin{align}
        \Delta\mathcal{T}^l_i & = \mathcal{T}^l(\omega_{i+1}) - \mathcal{T}^l(\omega_i) \nonumber \\
        & = (\omega_{i+1} + \omega_i)\T\frac{m}{2}(\omega_{i+1} - \omega_i) \nonumber\\
       \Delta\mathcal{T}_i^l & = (\omega_{i+1} + \omega_i)\T\frac{m}{2}(\omega_{i+1} - \omega_i).
\end{align}
The change in the value of the candidate Lyapunov function can be computed as,
\begin{align}
    \Delta V_i & = V_{i+1} - V_i = \Delta\mathcal{T}_i + k_p\Delta\mathcal{U}_i \nonumber\\
    & = \frac{1}{2}\left(\omega_{i+1} + \omega_i\right)\T\left(m(\omega_{i+1} - \omega_i) - k_phS_{L_i}(\hat{R}_i)\right).
\end{align}
Similarly, we obtain
\begin{align}
    \Delta V_{i+1} &= \frac{1}{2}\left(\omega_{i+2} + \omega_{i+1}\right)\T\left(m(\omega_{i+2} - \omega_{i+1}) \right. \nonumber \\ & \qquad \left. - k_phS_{L_{i+1}}(\hat{R}_{i+1})\right).
\end{align}
Substituting the value of $\omega_{i+2}$ from the filtering scheme presented in Proposition \ref{prop} we get
\begin{align}
    \Delta V_{i+1} & = \frac{1}{2}(\omega_{i+2} + \omega_{i+1})\T\Bigg\{ \exp{\left(-\frac{h}{2}(\hat{\Omega}_{i+2}+\hat{\Omega}_{i+1})^\times\right)} \Big \{ m(\omega_{i+1}+\omega_i) \Big. \Bigg. \nonumber
    \\ & \qquad \left. + hS_{L_{i+1}}(\hat{R}_{i+1}) - h\tau_{D_{i+1}} \right\} \Bigg. -2m\omega_{i+1} -  k_phS_{L_{i+1}}(\hat{R}_{i+1}) \Bigg\}.
\end{align}
Now, for $\Delta V$ to be negative definite for all $i$ we require
\begin{align}\label{eq:51}
    & \exp{\left(-\frac{h}{2}(\hat{\Omega}_{i+2}+\hat{\Omega}_{i+1})^\times\right)} \left \{ m(\omega_{i+1}+\omega_i) - \frac{h}{2}\tau_{D_{i+1}} +  \frac{h}{2}S_{L_{i+1}}(\hat{R}_{i+1}) \right\} \nonumber \\ & \qquad -2m\omega_{i+1} -  k_phS_{L_{i+1}}(\hat{R}_{i+1}) = -l(\omega_{i+2} + \omega_{i+1}),
\end{align}
where $l > 0, l \neq m$,
and $\Delta V_{i+1}$ simplifies to
\begin{equation}\label{eq:52}
    \Delta V_{i+1} = -\frac{l}{2}\left(\omega_{i+2} + \omega_{i+1}\right)\T\left(\omega_{i+2} + \omega_{i+1}\right).
\end{equation}
Substituting $\omega_{i+2}$ from the third equation presented in Proposition \ref{prop} into \eqref{eq:51},
\begin{align}
    & \exp{\left(-\frac{h}{2}(\hat{\Omega}_{i+2}+\hat{\Omega}_{i+1})^\times\right)} \left \{ m(\omega_{i+1}+\omega_i) - \frac{h}{2}\tau_{D_{i+1}} + \frac{h}{2}S_{L_{i+1}}(\hat{R}_{i+1}) \right\} -2m\omega_{i+1} \nonumber \\ & - k_phS_{L_{i+1}}(\hat{R}_{i+1}) = -\frac{l}{m}\exp{\left(-\frac{h}{2}(\hat{\Omega}_{i+1}+\hat{\Omega}_i)^\times\right)} \Bigg \{ m(\omega_{i+1}+\omega_i) \Bigg.  \nonumber
    \\ & \qquad\qquad\qquad\qquad\qquad\qquad\left. - \frac{h}{2}\tau_{D_{i+1}} +  \frac{h}{2}S_{L_{i+1}}(\hat{R}_{i+1}) \right\},
\end{align}
which further simplifies to,
\begin{align}
    & \frac{m+l}{m}\exp{\left(-\frac{h}{2}(\hat{\Omega}_{i+2}+\hat{\Omega}_{i+1})^\times\right)} \left \{ m(\omega_{i+1}+\omega_i) - \frac{h}{2}\tau_{D_{i+1}} +  \frac{h}{2}S_{L_{i+1}}(\hat{R}_{i+1}) \right\} \nonumber \\
    & \qquad = 2m\omega_{i+1} +  k_phS_{L_{i+1}}(\hat{R}_{i+1}),
\end{align}
which upon simple manipulations yields \eqref{eq:37}. It can be seen that after substituting \eqref{eq:37} into Proposition \ref{prop}, we obtain
\begin{equation}\label{eq:55}
    \omega_{i+2} = \frac{1}{m+l}\left[ (m-l)\omega_{i+1} + k_phS_{L_{i+1}}(\hat{R}_{i+1}) \right].
\end{equation}
\eqref{eq:55} can also be rewritten as
\begin{equation}\label{eq:56}
    \omega_{i+1} = \frac{1}{m+l}\left[ (m-l)\omega_i + k_phS_{L_i}(\hat{R}_i) \right],
\end{equation}
in terms of $\omega_i,\omega_{i+1}$ and $S_{L_i}(\hat{R}_i)$. From \eqref{eq:52}, $\Delta V_i$ can be written as
\begin{equation}
    \Delta V_i = \frac{-l}{2}(\omega_{i+1} + \omega_i)\T(\omega_{i+1} + \omega_i).
\end{equation}
We employ the discrete-time La-Salle invariance principle from \cite{lasalle1976stability} considering our domain ($\SO\times\bR^3$) to be a subset of $\bR^{12}$. We use Theorem 6.3 and Theorem 7.9  from Chapter-1 of \cite{lasalle1976stability}. For this we first compute $\mathscr{E} := \{(Q_i,\omega_i) \in \SO\times\bR^3 | \Delta V_i(Q_i,\omega_i) = 0\} = \{(Q_i,\omega_i) \in \SO\times\bR^3 \; | \; \omega_{i+1} + \omega_i = 0\}$. From \eqref{eq:39},  $\omega_{i+1} + \omega_i = 0$ implies that
\begin{equation}
    Q_{i+1} = Q_i.
\end{equation}
Also, from \eqref{eq:46} we have  $\Delta\mathcal{U} = 0$ whenever $\omega_{i+1} + \omega_i = 0$. This implies that the potential function, which is a Morse function according to lemma \ref{lemma:1}, is not changing and therefore has converged to one of its stationary points. Stationary points of the Morse function $\langle I-Q,K \rangle$ are characterised as the solutions of
\begin{equation}\label{eq:59}
    S_K(Q_i) = 0 \Rightarrow \text{vex}\left(KQ_i - Q_i\T K\right) = 0 \Rightarrow KQ_i = Q_i\T K.
\end{equation}
Now, $L_i = E_iW_i(\Tilde{U}_i^m)\T=E_iW_i(R_i\T E_i)\T= (E_iW_iE_i\T)R_i=KR_i$, which further gives us
\begin{align}\label{eq:61}
    \left(S_{L_i}(\hat{R}_i)\right)^\times & = L_i\T\hat{R}_i - \hat{R}_i\T  L_i \nonumber\\
    & = R_i\T K\hat{R}_i - \hat{R}_i\T KR_i.
\end{align}
Multiplying \eqref{eq:61} from the right hand side by $\hat{R}_i\T$  and from the left hand side by $\hat{R}_i$,
\begin{align}
    \hat{R}_i\left(S_{L_i}(\hat{R}_i)\right)^\times\hat{R}_i\T & = \hat{R}_iR_i\T K - KR_i\hat{R}_i\T \nonumber \\
    & = Q_i\T K - KQ_i.
\end{align}
At the critical points from \eqref{eq:59}, the right side of the above expression vanishes.
Therefore, as $\hat{R}_i$ is an orthogonal matrix, the following holds true at the critical points:
\begin{equation}
   \left(S_{L_i}(\hat{R}_i)\right)^\times = 0 \Rightarrow S_{L_i}(\hat{R}_i) = 0.
\end{equation}
Substituting this information in \eqref{eq:56} yields,
\begin{equation}
    \omega_{i+1} = \frac{1}{m+l} (m-l)\left(\omega_i \right).
\end{equation}
Now if, $\omega_{i+1} + \omega_i = 0$, we have
\begin{equation}
    \frac{2m}{m+l}\omega_i = 0 \Rightarrow \omega_i = \omega_{i+1} = 0.
\end{equation}
This leads to the conclusion that the set of estimation errors, $\mathscr{E} = \{(Q_i,\omega_i) \in \SO\times\bR^3 \; | \; Q_i \in C_Q, \omega_i = 0\}$, is the largest invariant set for the estimation error dynamics, and we obtain $\mathscr{M} = \mathscr{E} = \{(Q_i,\omega_i) \in \SO\times\bR^3 \; | \; Q_i \in C_Q, \omega_i = 0\}$. 
Therefore, we obtain the positive limit set as the set,
\begin{align}
    \mathscr{I} & := \mathscr{M} \cap V_i^{-1}(0) \nonumber \\ & = \{(Q,\omega) \in \SO\times\bR^3\; | \; Q \in C_Q, \omega = 0 \}.
\end{align}

In the absence of measurement errors, all the solutions of this filter converge asymptotically to the set $\mathscr{I}$. More specifically, the attitude estimation error converges to the set of critical points of $\langle\,I-Q,K\rangle$. The unique global minimum of this function is at $(Q,\omega) = (I,0)$ from lemma \ref{lemma:2}. Therefore, $(Q,\omega) = (I,0)$ is locally asymptotically stable. The remainder of this proof is similar to the last part of the proof of stability of the variational attitude estimator in \cite{izadi2014rigid}.

Consider the set,
\begin{equation}
    \mathscr{C} = \mathscr{I} \backslash (I,0)
\end{equation}
which consists of all the stationary states that the estimation errors may converge to,
besides the desired estimation error state $(I,0)$. Note that all states in the stable
manifold of a stationary state in $\mathscr{C}$ will converge to this stationary state. From
the properties of the critical points $Q^i \in C_Q \backslash (I)$ of $\mathit{\Phi}(\langle\,K,I-Q\rangle)$  given in lemma \ref{lemma:2}.  we see that the stationary points in $\mathscr{I}\backslash(I,0) = \{(Q^i,0) : Q^i \in C_Q \backslash (I)\}$ have stable manifolds whose dimensions depend on the index of $Q^i$. Since the angular velocity estimate error $\omega$ converges globally to the zero vector, the dimension of the stable manifold $\mathcal{M}_i^S$ of $(Q^i,0) \in \SO\times\bR^3$ is
\begin{equation}\label{eq:68}
    \text{dim}(\mathcal{M}_i^S) = 3 + (3 - \text{index of }\; Q^i) = 6 - \text{index of }\; Q^i.
\end{equation}

therefore, the stable manifolds of $(Q,\omega) = (Q^i,0)$ are three-dimensional, four
dimensional, or five-dimensional, depending on the index of $Q^i \in C_Q \backslash (I)$ according to \eqref{eq:68}. Moreover, the value of the Lyapunov function $V(Q_i,\omega_i)$ is non
decreasing (increasing when $(Q_i,\omega_i) \notin \mathscr{I}$) for trajectories on these manifolds when going backwards in time. This implies that the metric distance between error
states $(Q,\omega)$ along these trajectories on the stable manifolds $\mathcal{M}_i^S$ grows with the time separation between these states, and this property does not depend on the
choice of the metric on $\SO \times \bR^3$. Therefore, these stable manifolds are embedded (closed) sub-manifolds of $\SO \times \bR^3$ and so is their union. Clearly, all states starting in the complement of this union, converge to the stable equilibrium
$(Q,\omega) = (I,0)$; therefore the domain of attraction of this equilibrium is,
\begin{equation}
    DOA{(I,0)} = \SO\times\bR^3\backslash   \{\cup_{i=1}^3\mathcal{M}_i^S\}
\end{equation}
which is a dense open subset of $\SO\times\bR^3$.
\end{proof}

\section{Numerical Simulations}\label{sec:6}

This section presents numerical simulation results of the discrete-time estimator presented in the section \ref{sec:5}. The rigid body is assumed to have an initial attitude and angular velocity given by,
$$ R_0 = \text{expm}_{\SO}\left(\left( \frac{\pi}{4}\times\left[\frac{4}{7}, \; \frac{2}{7}, \; \frac{5}{7}\right]\T \right)^\times\right),$$ $$\text{and} \;\; \Omega_0 = \frac{\pi}{60}\times[-1.2, \; 2.1, \; -1.9]\T \; rad/s. $$

$W$ is selected based on the measured set of inertial vectors $E$ at each instant such that it satisfies lemma \ref{lemma:1}. Initially estimated states have the following initial estimation errors:
$$ Q_0 = \text{expm}_{\SO}\left(\left( \frac{\pi}{2.5}\times\left[\frac{4}{7}, \; \frac{2}{7}, \; \frac{5}{7}\right]\T \right)^\times\right),$$ $$\text{and} \;\; \omega_0 = [0.001, \; -0.002, \; 0.003]\T \; rad/s. $$

It has been assumed that there are at most 9 inertially known directions that are being measured by the sensors attached to the rigid body. The number of observed direction can vary randomly between 2 to 9 at each time instant. In the case where the number of observed directions is 2, the cross product of the two measurements is used as the third measurement. The standard rigid body dynamics are used to produce true states of the rigid body by applying sinusoidal forces. The observed directions in the body-fixed frame are simulated with the help of the aforementioned true states. The true quantities are disturbed by bounded, random noise with zero mean to simulate realistic measurements. Based on coarse attitude sensors like sun sensors and magnetometers, a random noise bounded in magnitude by $2.4^\circ$ is added to the matrix $U = R\T E$ to generate measured $U^m$. Similarly, a random noise bounded in magnitude by $0.97^\circ/s$, which is close to real noise levels of coarse rate gyros, is added $\Omega$ to generate measured $\Omega^m$.
\vspace{-0.2cm}
\begin{figure}[H]
	\centering
	\includegraphics[scale = 0.65]{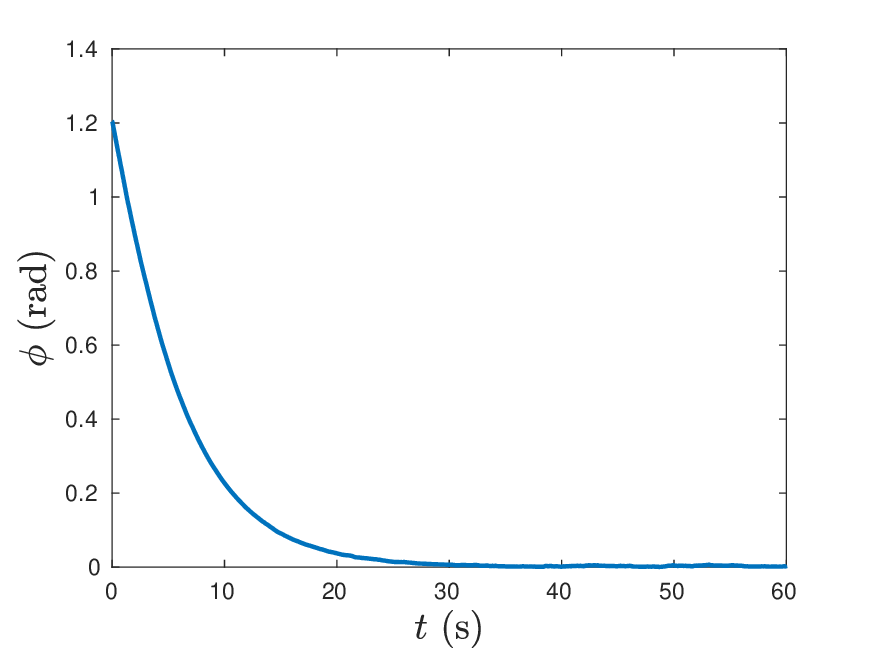}
	\caption{Case-1: Principle angle of the attitude estimation error}
	\label{Case_1_fig_2}
\end{figure}

\vspace{-0.5cm}
\subsection{Case-1 Simulation Results}
\begin{figure}
	\centering
	\includegraphics[scale = 0.65]{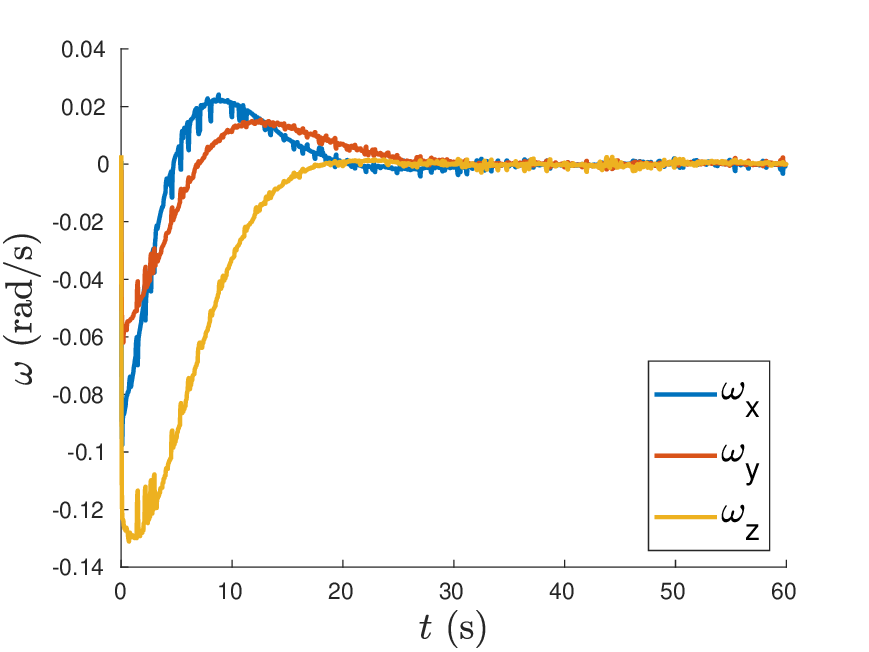}
	\caption{Case-1: Angular velocity estimation error}
	\label{Case_1_fig_1}
\end{figure}

The estimator is simulated over a time interval of $T$ = 60s, with a step-size of $h = 0.01 s$. The inertial scalar gain is $m = 1.5$ and the dissipation term is chosen to be $l = 0.3$. The difference of sampling rate between measurements of angular velocity and measurements inertial vectors in body-fixed frame is taken to be $n = 10$. Furthermore, the value of gain $k_p$ is chosen to be $k_p = 1$. The principle angle $\phi$ of the rigid body's attitude estimation error $Q$ is shown in the fig \ref{Case_1_fig_2}. Components of estimation error $\omega$ in the rigid body's angular velocity are shown in fig \ref{Case_1_fig_1}.

\subsection{Case-2 Simulation Results}

\begin{figure}
	\centering
	\includegraphics[scale = 0.65]{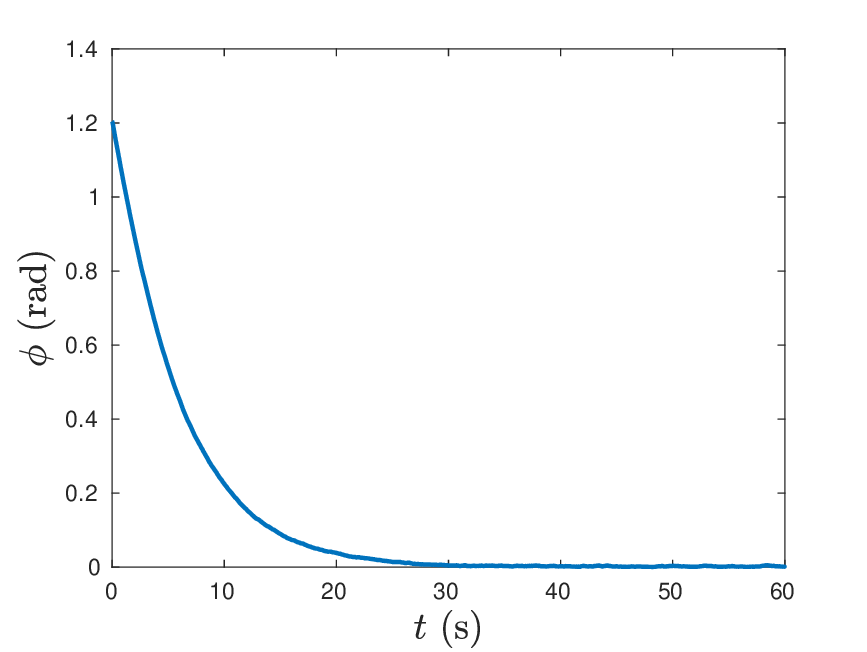}
	\caption{Case-2: Principle angle of the attitude estimation error}
	\label{Case_2_fig_2}
\end{figure}

\begin{figure}
	\centering
	\includegraphics[scale = 0.65]{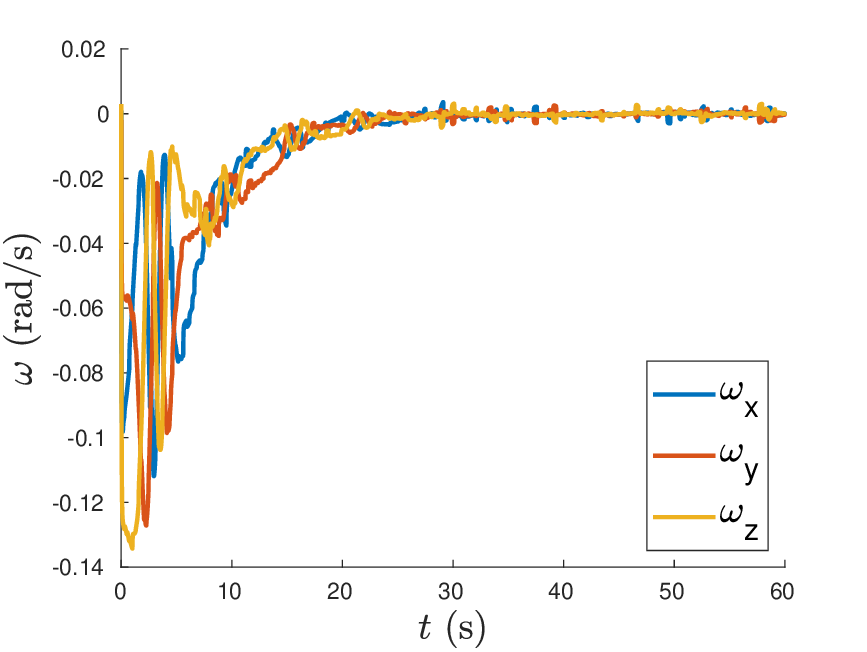}
	\caption{Case-2: Angular velocity estimation error}
	\label{Case_2_fig_1}
\end{figure}

The estimator is simulated over a time interval of $T$ = 60s, with a step-size of $h = 0.01 s$. The inertial scalar gain is $m = 1.5$ and the dissipation term is chosen to be $l = 0.3$. The difference of sampling rate between measurements of angular velocity and measurements inertial vectors in body-fixed frame is to varying randomly but bounded between 10 and 30. Furthermore, the value of gain $k_p$ is chosen to be $k_p = 1$. The principle angle $\phi$ of the rigid body's attitude estimation error $Q$ is shown in the fig \ref{Case_2_fig_2}. Components of estimation error $\omega$ in the rigid body's angular velocity are shown in fig \ref{Case_2_fig_1}.
\vspace*{-2mm}

\subsection{Case-3 Simulation Results}

\begin{figure}
	\centering
	\includegraphics[scale = 0.65]{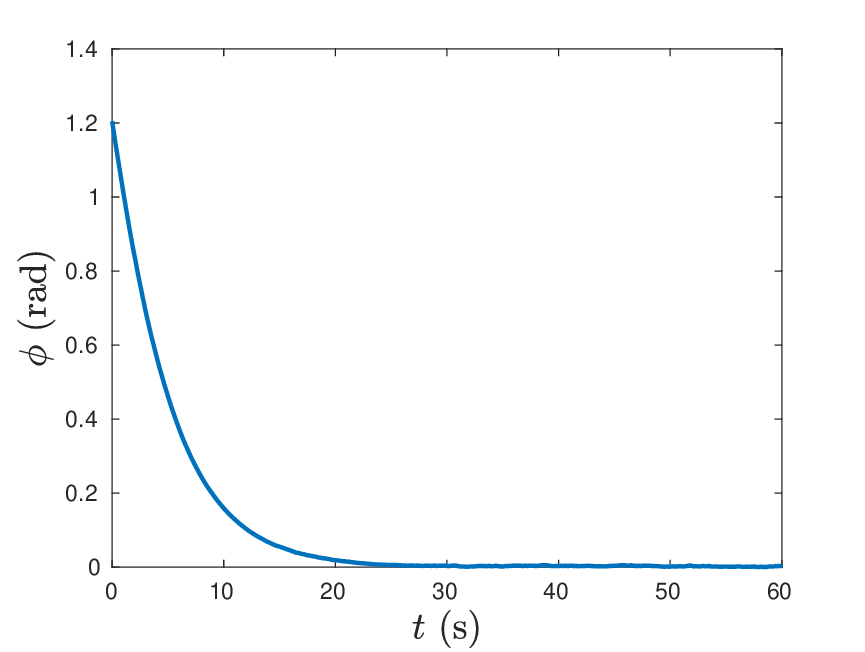}
	\caption{Case-3: Principle angle of the attitude estimation error}
	\label{Case_3_fig_2}
\end{figure}

\begin{figure}
	\centering
	\includegraphics[scale = 0.65]{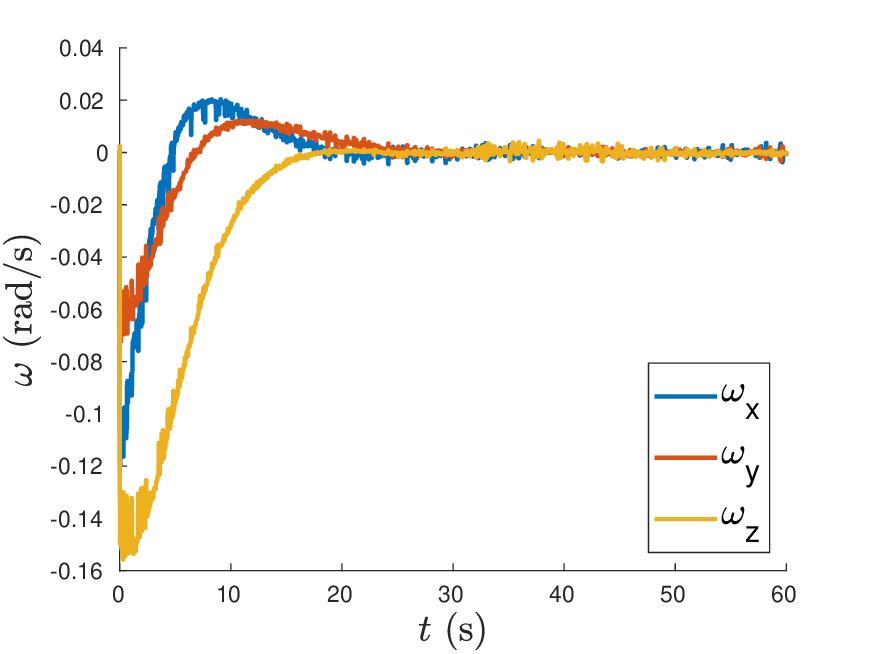}
	\caption{Case-3: Angular velocity estimation error}
	\label{Case_3_fig_1}
\end{figure}

The estimator is simulated over a time interval of $T$ = 60s, with a step-size of $h = 0.008 s$. The inertial scalar gain is $m = 2.5$ and the dissipation term is chosen to be $l = 0.5$. The assumption is that the angular velocity measurements are available after each time interval of 0.008s and inertial vector measurements are available after each time interval of 0.05s. Furthermore, the value of gain $k_p$ is chosen to be $k_p = 10$. The principle angle $\phi$ of the rigid body's attitude estimation error $Q$ is shown in the fig \ref{Case_3_fig_2}. Components of estimation error $\omega$ in the rigid body's angular velocity are shown in fig \ref{Case_3_fig_1}. \\

\vspace*{-2mm}



{\color{black}

\subsection{Performance of the estimator in the presence of large initial estimaition errors}

We have discussed the performance of the estimator in various scenarios but with the fixed values of initial estimation errors. It remains to see whether the estimator perform good in case of large initial estimation errors. For this part, we restrict our attention to the Case-2 i.e., when the difference of sampling rate between measurement of angular velocity and measurement of inertial vectors is time varying and bounded between 10 and 30. We check the performance of the estimator for three different values of initial estimation errors in angular velocity and attitude. The results of the same are plotted in fig \ref{fig:1}.

On comparing fig \ref{fig:1a} and fig \ref{fig:1b}, when the initial estimation error in principle angle of attitude estimation is increased to about 2.5 times and estimation error in angular velocity is increased to 10 times, convergence takes about 10s more time. However, on comparing fig \ref{fig:1a} and fig \ref{fig:1c}, it can be see that if the initial attitude estimation error is fixed then change in initial angular velocity estimation error(100 times the original values) has no effect on the convergence time. Therefore, it can be concluded that increasing the initial attitude estimation error can increase the convergence time but increase in initial angular velocity estimation error has no effect on the convergence time.

\begin{figure}
     \centering
     \begin{subfigure}{\textwidth}
         \centering
         \includegraphics[scale=0.3]{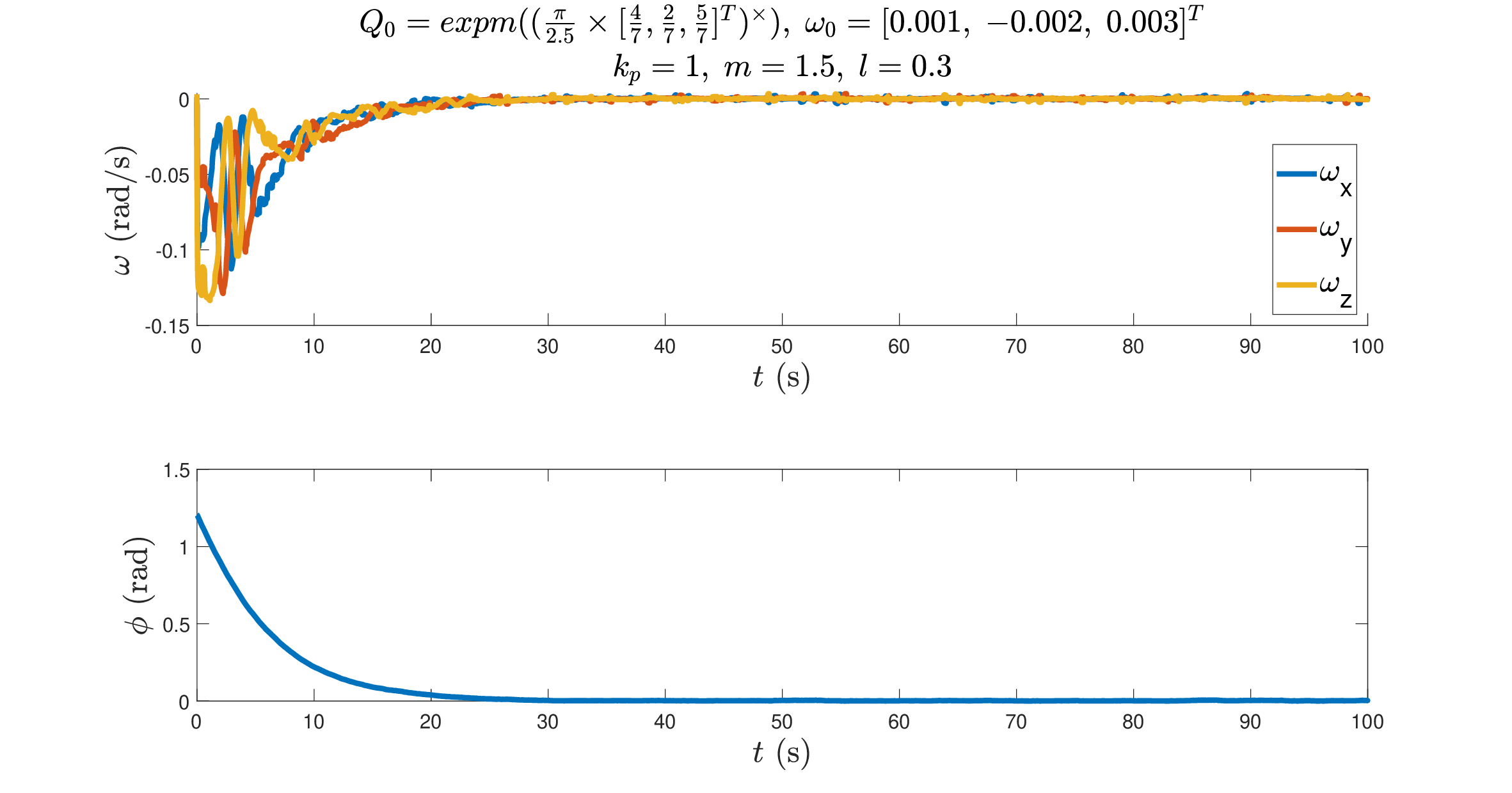}
         \vspace{-0.3cm}
         \caption{}
         \label{fig:1a}
     \end{subfigure}
     \begin{subfigure}{\textwidth}
         \centering
         \includegraphics[scale=0.3]{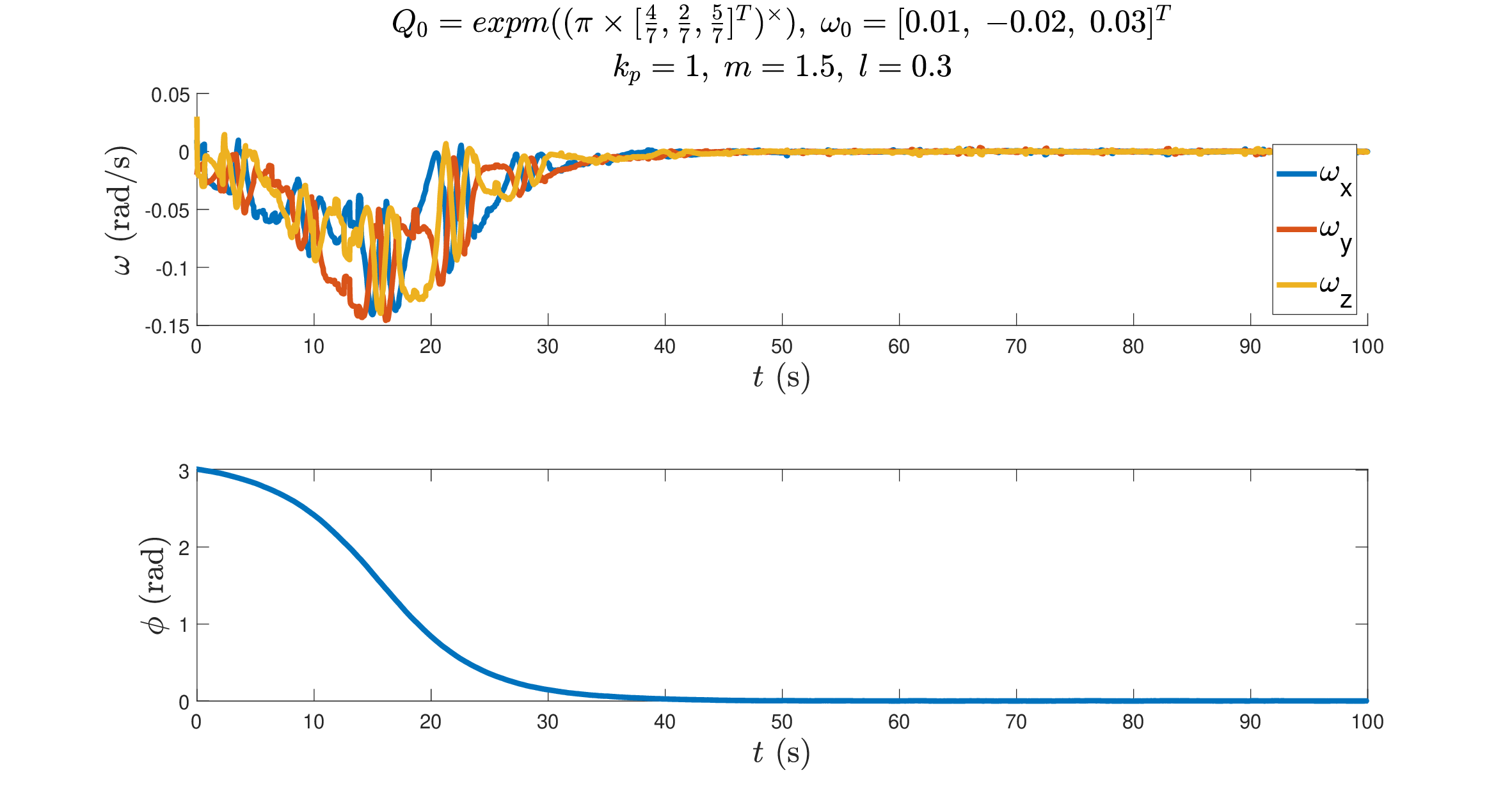}
         \vspace{-0.3cm}
         \caption{}
         \label{fig:1b}
     \end{subfigure}
     \hfill
     \begin{subfigure}{\textwidth}
         \centering
         \includegraphics[scale=0.3]{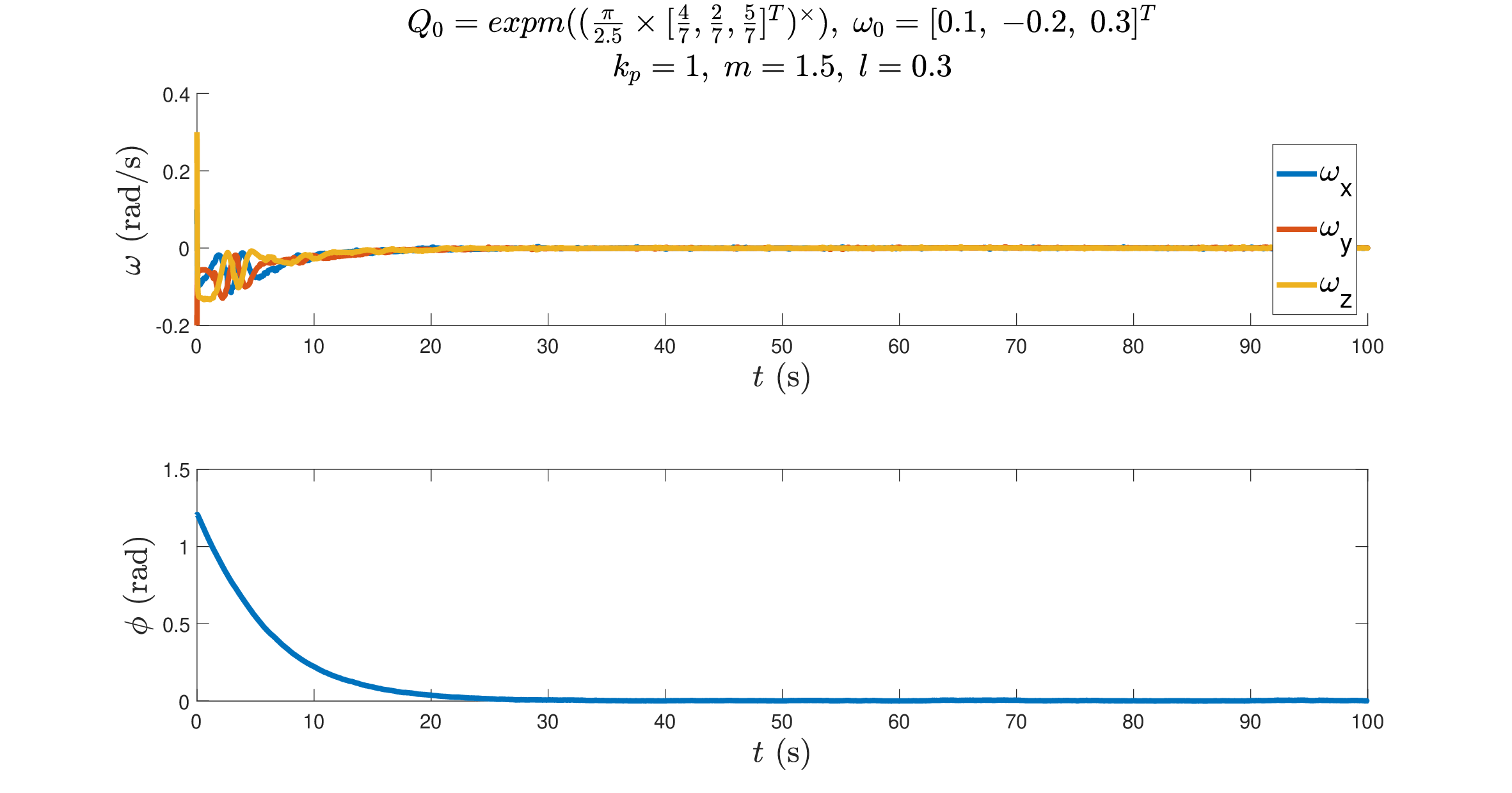}
         \vspace{-0.3cm}
         \caption{}
         \label{fig:1c}
     \end{subfigure}
        \caption{Performance of the estimator for different values of initial estimation errors}
        \label{fig:1}
\end{figure}

\subsection{Effect of gain on convergence time}
In this section, we discuss how convergence time can be decreased by increasing the value of gain. Same as before, we restrict our attention to the Case-2 i.e., when the difference of sampling rate between measurement of angular velocity and measurement of inertial vectors is time varying and bounded between 10 and 30. We consider very high errors in initial estimation errors and plot the performance of the estimator for three different values of gain $k_p$ in fig \ref{fig:6}. As it can be seen, increasing the value of gain significantly improves the convergence time.

\begin{figure}
     \centering
     \begin{subfigure}{\textwidth}
         \centering
         \includegraphics[scale=0.3]{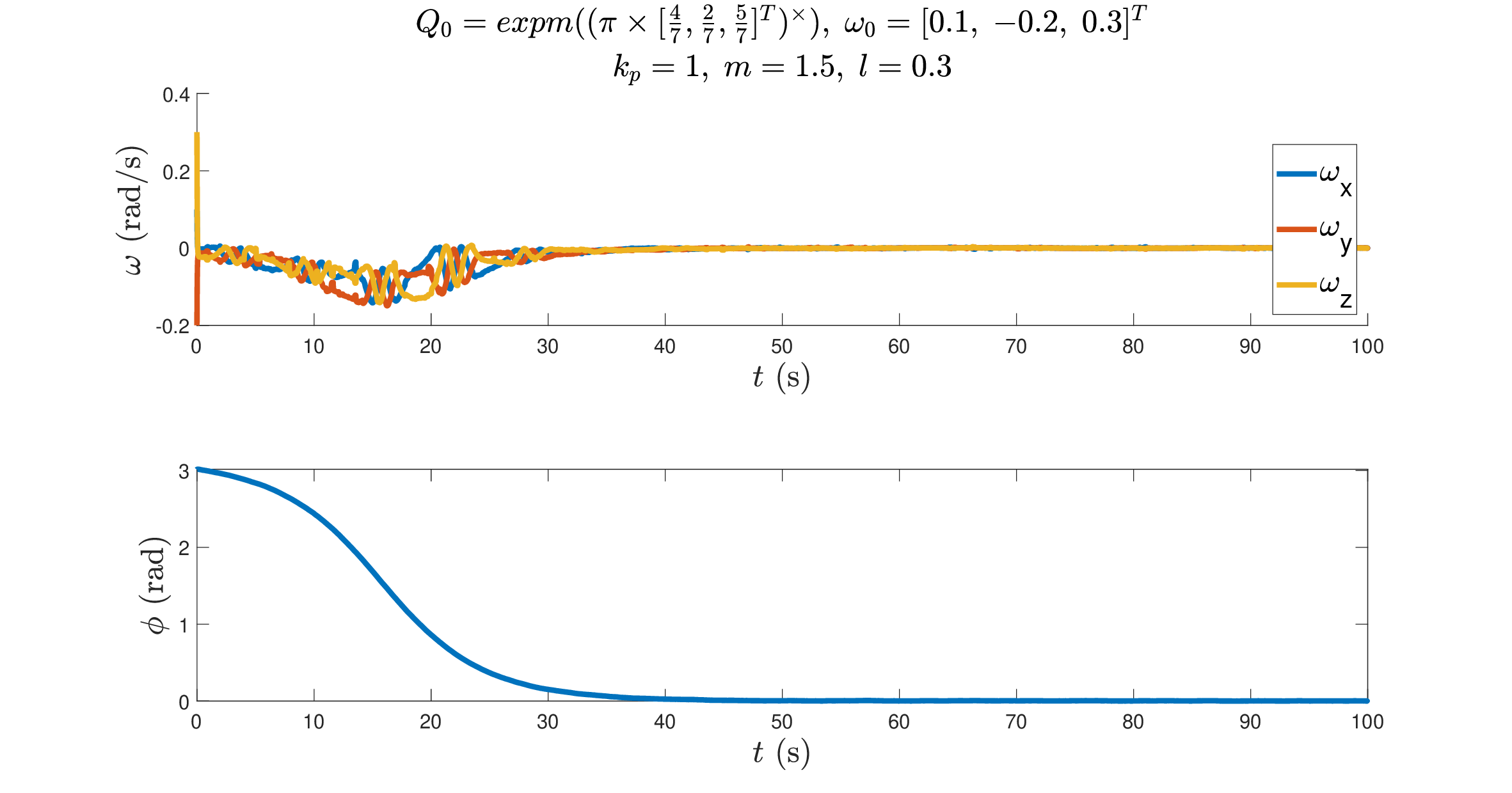}
         \vspace{-0.3cm}
         \caption{}
         \label{fig:4a}
     \end{subfigure}
     \hfill
     \begin{subfigure}{\textwidth}
         \centering
         \includegraphics[scale=0.3]{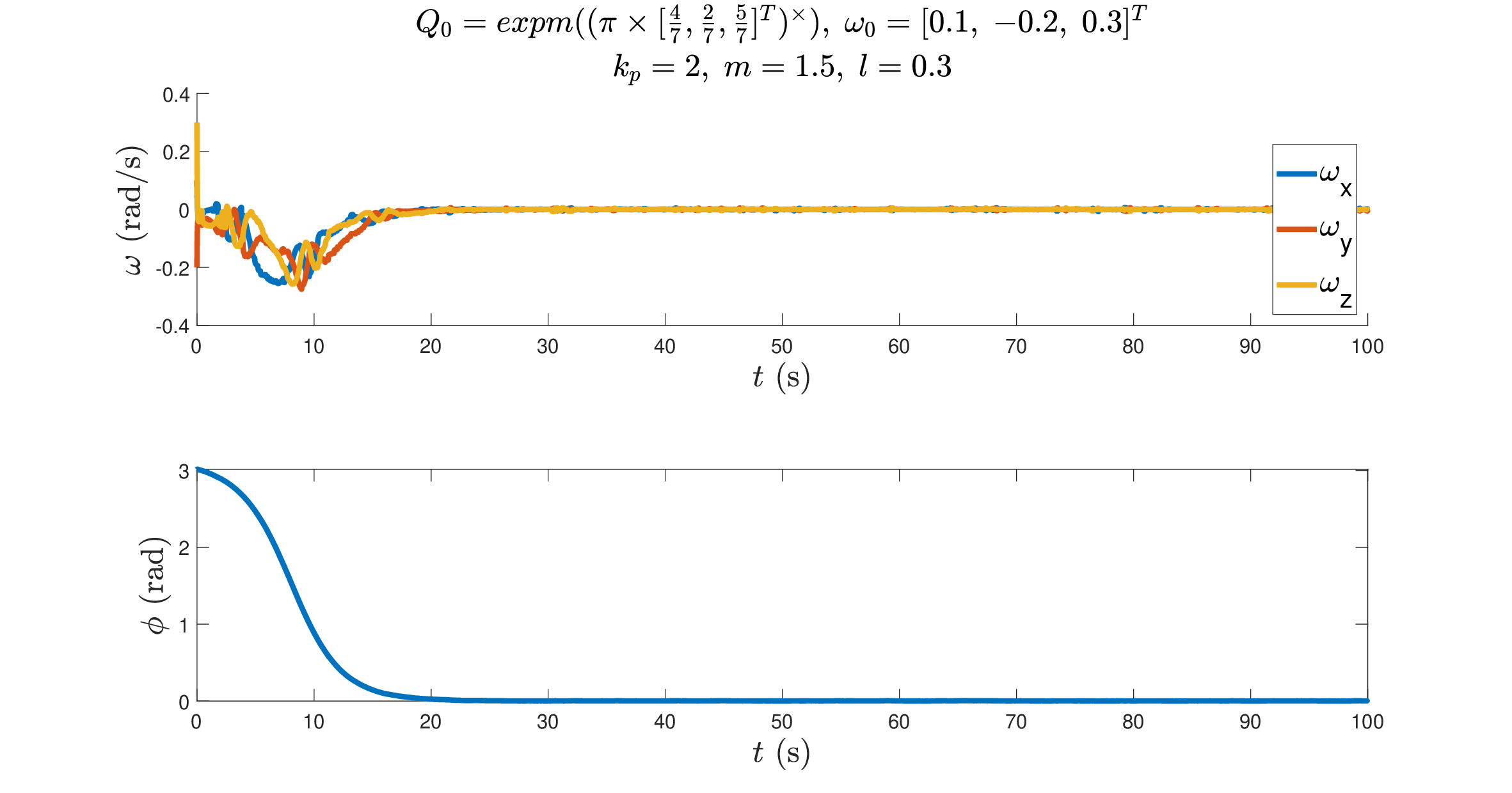}
         \vspace{-0.3cm}
         \caption{}
         \label{fig:5b}
     \end{subfigure}
     \hfill
     \begin{subfigure}{\textwidth}
         \centering
         \includegraphics[scale=0.3]{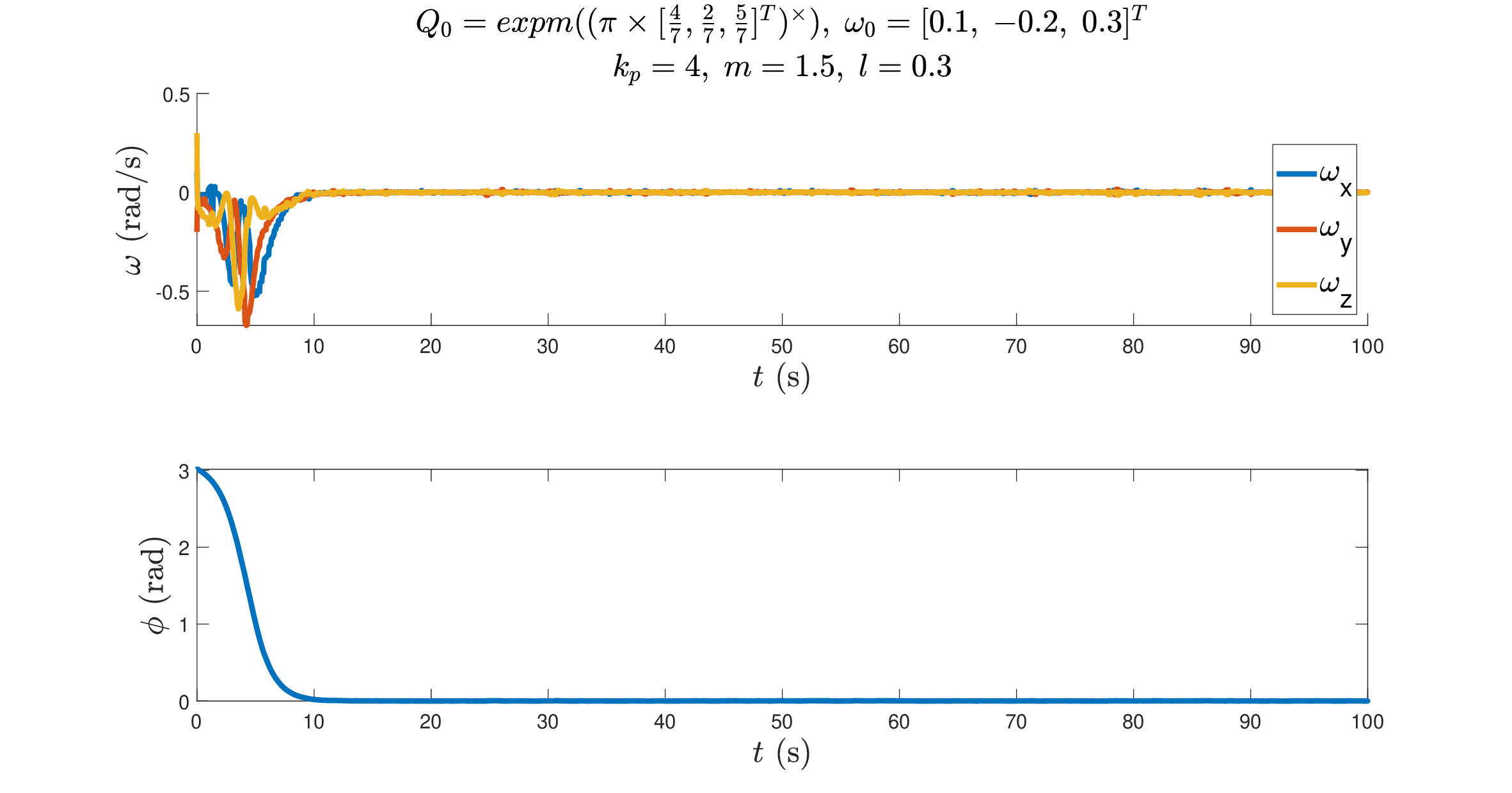}
         \vspace{-0.3cm}
         \caption{}
         \label{fig:6c}
     \end{subfigure}
        \caption{Performance of the estimator for different values of gain $k_p$}
        \label{fig:6}
\end{figure}

}

\section{Conclusion}\label{sec:7}

We develop a geometric attitude and angular velocity estimation scheme using the discrete-time Lagrange-D'Alembert principle followed by discrete-time Lyapunov stability analysis in the presence of multi-rate measurements. The attitude determination problem from two or more vector measurements in the body-fixed frame is formulated as Wahba's optimization problem. To overcome the multi-rate challenge, a discrete-time model for attitude kinematics is used to propagate the inertial vector measurements forward in time. The filtering scheme is obtained with the aid of appropriate discrete-time Lagrangian and Lyapunov functions consisting of Wahba's cost function as an artificial potential term and a kinetic energy-like term that is quadratic in the angular velocity estimation error. As it can be observed, the Lyapunov function is not constructed from the same artificial potential and kinetic energy terms that are used for constructing the Lagrangian. There are mainly two reasons behind this; 1) the filtering scheme obtained by applying the discrete Lagrange-d'Alembert principle is implicit in nature and therefore it can increase computational load and runtime making it difficult to use for real-time applications. Therefore, an explicit filtering scheme is more desirable, 2) we also need the filtering scheme to be asymptotically stable. A Lyapunov function, different from system energy constructed appropriately, helps us meet both the requirements. The explicit filtering scheme obtained after the Lyapunov analysis was proven to be asymptotically stable in the absence of measurement noise and the domain of convergence is proven to be almost global. Numerical simulations were carried out with realistic measurement data corrupted by bounded noise. Numerical simulations verified that the estimated states converge to a bounded neighborhood of $(I,0)$. Furthermore, the rate of convergence of the estimated states to the real state can be controlled by choosing appropriate gains. Future endeavors are towards obtaining a discrete-time optimal attitude estimator in the presence of multi-rate measurements when there a constant or slowly time-varying bias in the measurements of angular velocity while also obtaining a bound on the state estimation errors when there is measurement noise in the inertial vector measurements and the angular velocity measurements.











\begin{thebibliography}{10}

\bibitem{BERKANE2019415}
\newblock S.~Berkane and A.~Tayebi,
\newblock Attitude estimation with intermittent measurements,
\newblock \emph{Automatica}, \textbf{105} (2019), 415 -- 421.

\bibitem{bhatt2020rigid}
\newblock M.~Bhatt, S.~Sukumar and A.~K. Sanyal,
\newblock Rigid body geometric attitude estimator using multi-rate sensors,
\newblock in \emph{2020 59th IEEE Conference on Decision and Control (CDC)},
\newblock IEEE, 2020,
\newblock 1511--1516.

\bibitem{bhatt2020optimal}
\newblock M.~Bhatt, S.~Sukumar and A.~K. Sanyal,
\newblock Optimal multi-rate rigid body attitude estimation based on Lagrange-d'Alembert principle,
\newblock 2020

\bibitem{black1964passive}
\newblock H.~D. Black,
\newblock A passive system for determining the attitude of a satellite,
\newblock \emph{AIAA journal}, \textbf{2} (1964), 1350--1351.

\bibitem{blbook}
\newblock A.~M. Bloch, J.~Baillieul, P.~Crouch and J.~Marsden,
\newblock \emph{Nonholonomic Mechanics and Control},
\newblock 2nd edition,
\newblock no.~24 in Interdisciplinary Texts in Mathematics, Springer-Verlag,
  2015.

\bibitem{crassidis2007survey}
\newblock J.~L. Crassidis, F.~L. Markley and Y.~Cheng,
\newblock Survey of nonlinear attitude estimation methods,
\newblock \emph{Journal of guidance, control, and dynamics}, \textbf{30}
  (2007), 12--28.

\bibitem{davenport1968vector}
\newblock P.~B. Davenport,
\newblock \emph{A vector approach to the algebra of rotations with
  applications}, vol. 4696,
\newblock National Aeronautics and Space Administration, 1968.

\bibitem{goldstein1980classical}
\newblock H.~Goldstein and C.~Poole,
\newblock \emph{Classical Mechanics},
\newblock Addison-Wesley series in physics, Addison-Wesley Publishing Company,
  1980,
\newblock \urlprefix\url{https://books.google.co.in/books?id=9M8QAQAAIAAJ}.

\bibitem{greenwood1997classical}
\newblock D.~Greenwood,
\newblock \emph{Classical Dynamics},
\newblock Dover books on mathematics, Dover Publications, 1997,
\newblock \urlprefix\url{https://books.google.co.in/books?id=x7rj83I98yMC}.

\bibitem{izadi2014rigid}
\newblock M.~Izadi and A.~K. Sanyal,
\newblock Rigid body attitude estimation based on the lagrange--d’alembert
  principle,
\newblock \emph{Automatica}, \textbf{50} (2014), 2570--2577.

\bibitem{khosravian2015recursive}
\newblock A.~Khosravian, J.~Trumpf, R.~Mahony and T.~Hamel,
\newblock Recursive attitude estimation in the presence of multi-rate and
  multi-delay vector measurements,
\newblock in \emph{2015 American Control Conference (ACC)},
\newblock IEEE, 2015,
\newblock 3199--3205.

\bibitem{lasalle1976stability}
\newblock J.~LaSalle,
\newblock \emph{The Stability of Dynamical Systems}, vol.~25,
\newblock SIAM, 1976.

\bibitem{madinehi2013rigid}
\newblock N.~Madinehi,
\newblock Rigid body attitude estimation: An overview and comparative study,
\newblock \emph{Electronic Thesis and Dissertation Repository}.

\bibitem{mahony2008nonlinear}
\newblock R.~Mahony, T.~Hamel and J.-M. Pflimlin,
\newblock Nonlinear complementary filters on the special orthogonal group,
\newblock \emph{IEEE Transactions on automatic control}, \textbf{53} (2008),
  1203--1218.

\bibitem{markley1988attitude}
\newblock F.~L. Markley,
\newblock Attitude determination using vector observations and the singular
  value decomposition,
\newblock \emph{Journal of the Astronautical Sciences}, \textbf{36} (1988),
  245--258.

\bibitem{markley1993attitude}
\newblock F.~L. Markley,
\newblock Attitude determination using vector observations: A fast optimal
  matrix algorithm.

\bibitem{marsden2001discrete}
\newblock J.~E. Marsden and M.~West,
\newblock Discrete mechanics and variational integrators,
\newblock \emph{Acta Numerica}, \textbf{10} (2001), 357--514.

\bibitem{mortari1997esoq}
\newblock D.~Mortari,
\newblock {ESOQ}: A closed-form solution to the {Wahba} problem,
\newblock \emph{Journal of the Astronautical Sciences}, \textbf{45} (1997),
  195--204.

\bibitem{psiaki2012numerical}
\newblock M.~L. Psiaki and J.~C. Hinks,
\newblock Numerical solution of a generalized wahba problem for a spinning
  spacecraft,
\newblock \emph{Journal of Guidance, Control, and Dynamics}, \textbf{35}
  (2012), 764--773.

\bibitem{sany_acc06}
\newblock A.~K. {Sanyal},
\newblock Optimal attitude estimation and filtering without using local
  coordinates part i: Uncontrolled and deterministic attitude dynamics,
\newblock in \emph{2006 American Control Conference}, 2006,
\newblock 5734--5739.

\bibitem{sanyal2012attitude}
\newblock A.~K. Sanyal and N.~Nordkvist,
\newblock Attitude state estimation with multirate measurements for almost
  global attitude feedback tracking,
\newblock \emph{Journal of Guidance, Control, and Dynamics}, \textbf{35}
  (2012), 868--880.

\bibitem{shuster1981three}
\newblock M.~D. Shuster and S.~D. Oh,
\newblock Three-axis attitude determination from vector observations,
\newblock \emph{Journal of guidance and Control}, \textbf{4} (1981), 70--77.

\bibitem{valpiani2008nonlinear}
\newblock J.~M. Valpiani and P.~L. Palmer,
\newblock Nonlinear geometric estimation for satellite attitude,
\newblock \emph{Journal of guidance, control, and dynamics}, \textbf{31}
  (2008), 835--848.

\bibitem{vasconcelos2007landmark}
\newblock J.~F. Vasconcelos, R.~Cunha, C.~Silvestre and P.~Oliveira,
\newblock Landmark based nonlinear observer for rigid body attitude and
  position estimation,
\newblock in \emph{2007 46th IEEE Conference on Decision and Control},
\newblock IEEE, 2007,
\newblock 1033--1038.

\bibitem{vasconcelos2008nonlinear}
\newblock J.~F. Vasconcelos, C.~Silvestre and P.~Oliveira,
\newblock A nonlinear observer for rigid body attitude estimation using vector
  observations,
\newblock \emph{IFAC Proceedings Volumes}, \textbf{41} (2008), 8599--8604.

\bibitem{wahba1965least}
\newblock G.~Wahba,
\newblock A least squares estimate of satellite attitude,
\newblock \emph{SIAM review}, \textbf{7} (1965), 409--409.

\bibitem{zamani2010near}
\newblock M.~Zamani, J.~Trumpf and R.~Mahony,
\newblock Near-optimal deterministic attitude filtering,
\newblock in \emph{49th IEEE Conference on Decision and Control (CDC)},
\newblock IEEE, 2010,
\newblock 6511--6516.

\bibitem{zamani2013minimum}
\newblock M.~Zamani, J.~Trumpf and R.~Mahony,
\newblock Minimum-energy filtering for attitude estimation,
\newblock \emph{IEEE Transactions on Automatic Control}, \textbf{58} (2013),
  2917--2921.

\end{thebibliography}

\providecommand{\href}[2]{#2}
\providecommand{\arxiv}[1]{\href{http://arxiv.org/abs/#1}{arXiv:#1}}
\providecommand{\url}[1]{\texttt{#1}}
\providecommand{\urlprefix}{URL }

\medskip
Received xxxx 20xx; revised xxxx 20xx.
\medskip

\end{document}